\documentclass[12pt]{nature}

\usepackage{footmisc}
\usepackage{fixfoot}
\usepackage{booktabs}
\usepackage{url}
\usepackage[labelformat=simple]{subfig}
\usepackage[nomarkers, nolists,figuresonly]{endfloat}
\usepackage[]{graphicx}
\usepackage{bibunits}
\usepackage{dcolumn}
\usepackage{bm}
\usepackage{times}
\usepackage{amsmath}
\usepackage{siunitx}
\usepackage[font={footnotesize},labelfont={bf,sf},textfont={sf}]{caption}
\usepackage{psfrag}
\usepackage{booktabs}
\usepackage{array}
\usepackage{xr}
\usepackage[left]{lineno}
\usepackage{floatrow}

\makeatletter

\let\saved@includegraphics\includegraphics
\AtBeginDocument{\let\includegraphics\saved@includegraphics}

\DeclareCaptionLabelSeparator{bar}{ \vrule width 1pt depth 1pt height 8pt \kern 3pt }
\newcommand{\ssection}[1]{\section[#1]{\large \textsf {#1}}}
\newcommand{\tabincell}[2]{\begin{tabular}{@{}#1@{}}#2\end{tabular}}
\renewenvironment*{figure}{\@float{figure}}{\end@float}
\newcommand{\bcaption}[2]{\caption{\textbf{#1} #2}}

\title{A stacked prism lens concept for next generation hard X-ray telescopes}

\author{Wujun Mi$^{1*}$, Peter Nillius$^{1}$, Mark Pearce$^{1,2}$ \& Mats Danielsson$^1$ }

\selectfont

\begin{document}

\rm
\maketitle

\begin{affiliations}
\item KTH Royal Institute of Technology, Department of Physics, 106 91, Stockholm, Sweden.$^2$ The Oskar Klein Centre for Cosmoparticle Physics, AlbaNova University Centre, 106 91, Stockholm, Sweden. *e-mail: wujun.mi@mi.physics.kth.se
\end{affiliations}

\begin{abstract}
\textbf{
Effective collecting area, angular resolution, field of view and energy response are fundamental attributes of X-ray telescopes. The performance of state-of-the-art telescopes is currently restricted by Wolter optics, especially for hard X-rays. In this paper, we report the development of a new approach - the Stacked Prism Lens, which is lightweight, modular and has the potential for a significant improvement in effective area, while retaining high angular resolution. The proposed optics is built by stacking discs embedded with prismatic rings, created with photoresist by focused UV lithography. We demonstrate the SPL approach using a prototype lens which was manufactured and characterized at a synchrotron radiation facility. The design of a potential satellite-borne X-ray telescope is outlined and the performance is compared to contemporary missions.}
\end{abstract}

\vspace{1cm}

Since the first orbiting X-ray telescope, the Einstein Observatory, was launched in 1978, focusing X-ray telescopes (XRTs) have provided new knowledge on the universe by observing remote objects in the X-ray spectrum\cite{karouzos2017x,gorenstein2010focusing}. The performance of XRTs is mainly determined by the optics. State-of-the-art focusing XRTs rely on Wolter optics, for which X-rays that are nearly parallel to the nested mirrors are collected by total external reflection. This has been successfully employed in several telescopes (e.g. Chandra\cite{weisskopf2000chandra}, XMM-Newton\cite{jansen2001xmm}, Swift\cite{gehrels2004swift} and the planned ATHENA mission\cite{nandra2013hot}), for X-rays in the energy range of 0.1 to 10 keV with high values of efficiency, angular resolution and sensitivity. However, for X-ray energies higher than 10 keV, using the same technique compromises spatial resolution and efficiency since the grazing angle quickly decreases with energy, making the focal length of the system impractically long and the field of view (FoV) very small. Moreover, the efficiency of the mirrors decreases, and nesting becomes more difficult. To mitigate this issue, it is possible to use multi-layered coatings and Bragg-reflection from depth-graded multi-layers to increase the grazing angle in the hard X-ray energy range\cite{koglin2009nustar, koglin2003development}. Current hard X-ray focusing telescopes such as NuSTAR\cite{harrison2013nuclear1} and Astro-H\cite{takahashi2012astro,awaki2017hitomi} are designed this way. Although this provides a significant improvement in performance for hard X-rays, the long focal length is challenging when designing missions, and the small effective collecting area and narrow FoV limits the scientific return from missions. In addition, the angular resolution is severely influenced by figure error, surface roughness and assembly precision\cite{gorenstein2010focusing}.

At present, refractive\cite{Cederstroem2000,snigirev1996compound} and diffractive transmissive X-ray optics\cite{chang2014ultra} have been used in a wide range of applications\cite{sakdinawat2010nanoscale, chao2005soft, di1999high}, such as biology\cite{schneider2010three}, chemistry\cite{shapiro2014chemical} and lithography\cite{leontowich2011zone,larciprete2002direct}. Several novel zone plates\cite{chang2014ultra}, such as multilayer zone plates\cite{keskinbora2014multilayer, kang2008focusing, huang201311, kang2006nanometer}, interlaced zone plates\cite{mohacsi2017interlaced} and stacked zone plates\cite{maser2002near}, as well as refractive optics, such as commercially available compound refractive X-ray lenses\cite{snigirev1996compound}, have been developed for hard X-ray imaging. The excellent resolution of these lenses makes them very competitive, but their limited apertures, low efficiency and fabrication difficulties remain obstacles for any direct application in hard X-ray astronomy. Various XRT concepts based on diffractive and refractive optics\cite{skinner2001diffractive, refId0, skinner2008milli, braig2010fresnel, braig2010multiband} have been proposed and investigated for many years, with milli-arcsecond resolution expected. However, nearly all of them work only for X-ray energies \textless 11 keV and the focal length, up to thousands of kilometers between optics and detector spacecrafts, makes it impractical to turn these ideas into mature systems.

In this paper, we demonstrate a new concept in point-focusing X-ray optics, the Stacked Prism Lens (SPL). These lenses are based on SU-8 polymer photoresist and fabricated by focused UV lithography\cite{mi2016fabrication} using a conventional mask aligner, which is easily adopted for mass production. The performance as compared to current X-ray optics is considerably improved and most importantly, our new method allows arrays of X-ray lenses to form a scaleable modular assembly which has the potential to remove today's hard limits on effective area and FoV for hard X-ray focusing telescopes, while still maintaining an excellent angular resolution.

\ssection{Results}
\subsection{{Pricinple of Stacked Prism Lenses.}}
\begin{figure*}[ht!!]
 \centering
  \sidesubfloat[labelformat=simple]{
  	 \includegraphics[width = 5.5cm]{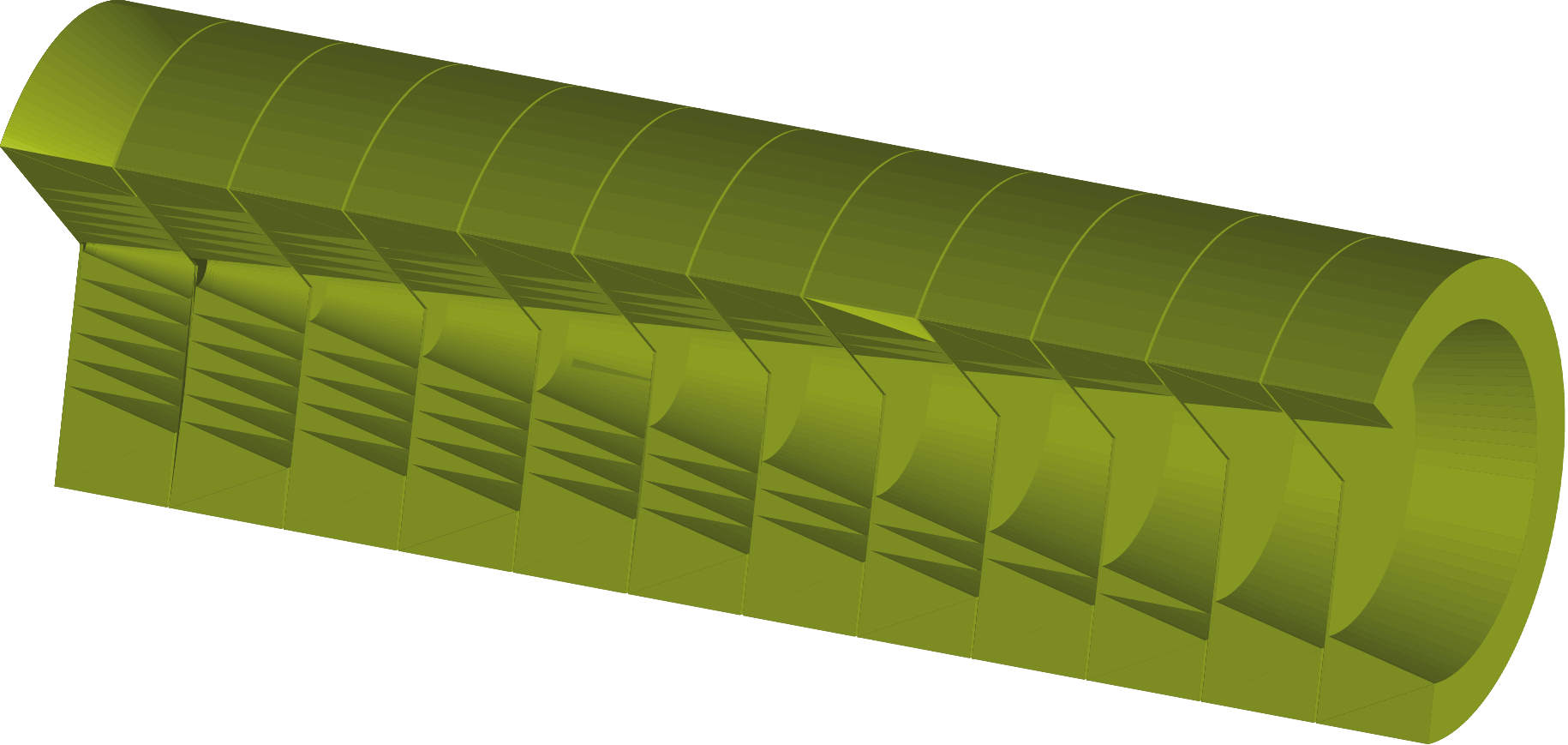} 
  	 \label{F:stacklens1}}
	 \hspace{1cm}
  \sidesubfloat[]{
  	\includegraphics[width = 4.4cm]{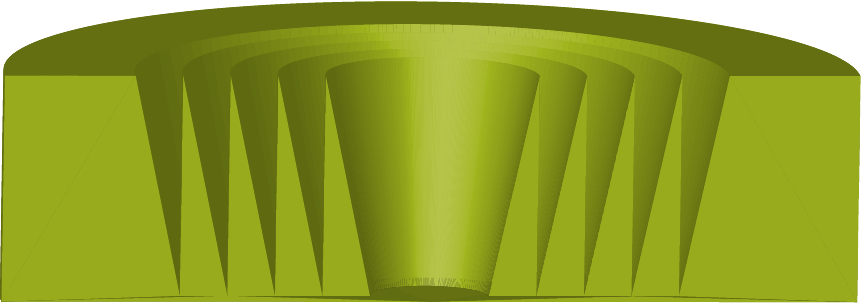}
	 \label{F:stacklens2}}	 
	 
   \sidesubfloat[]{
  	 \includegraphics[width = 12cm]{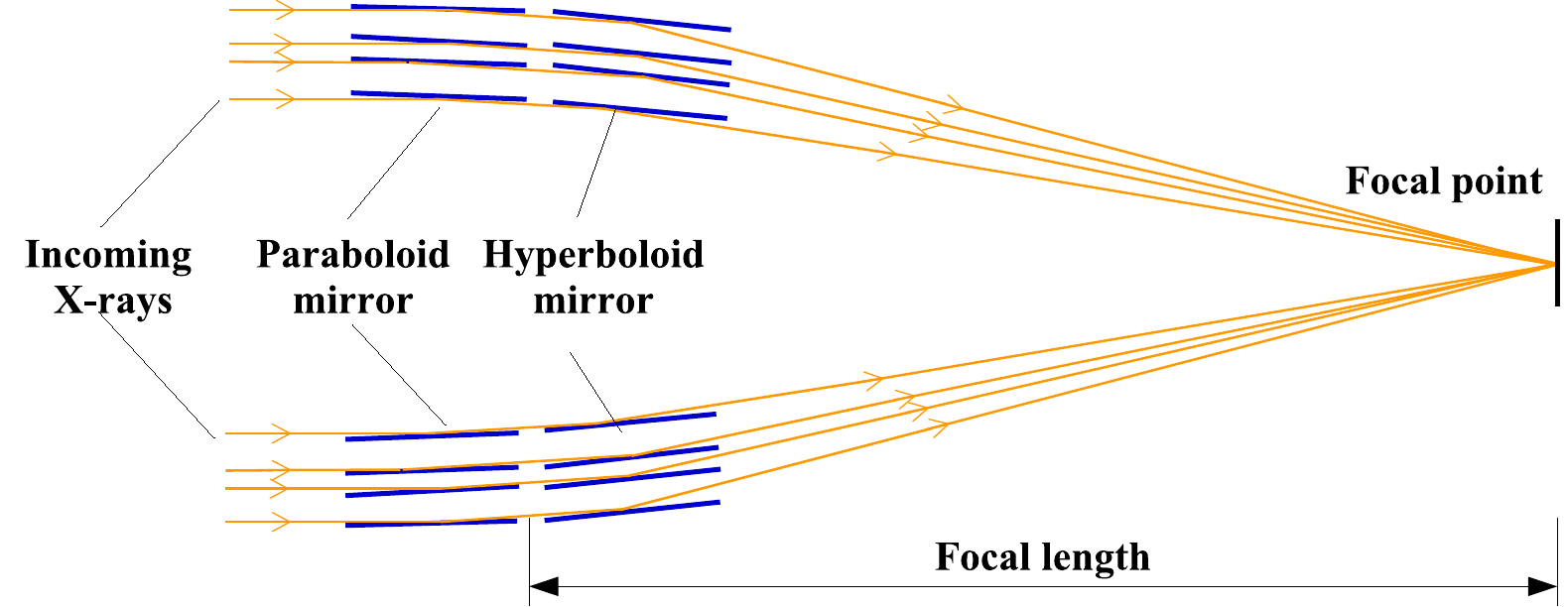}
  	 \label{F:wolterscheme}}
	 \hspace{1.7cm} 
   \sidesubfloat[]{
  	\includegraphics[width = 12cm]{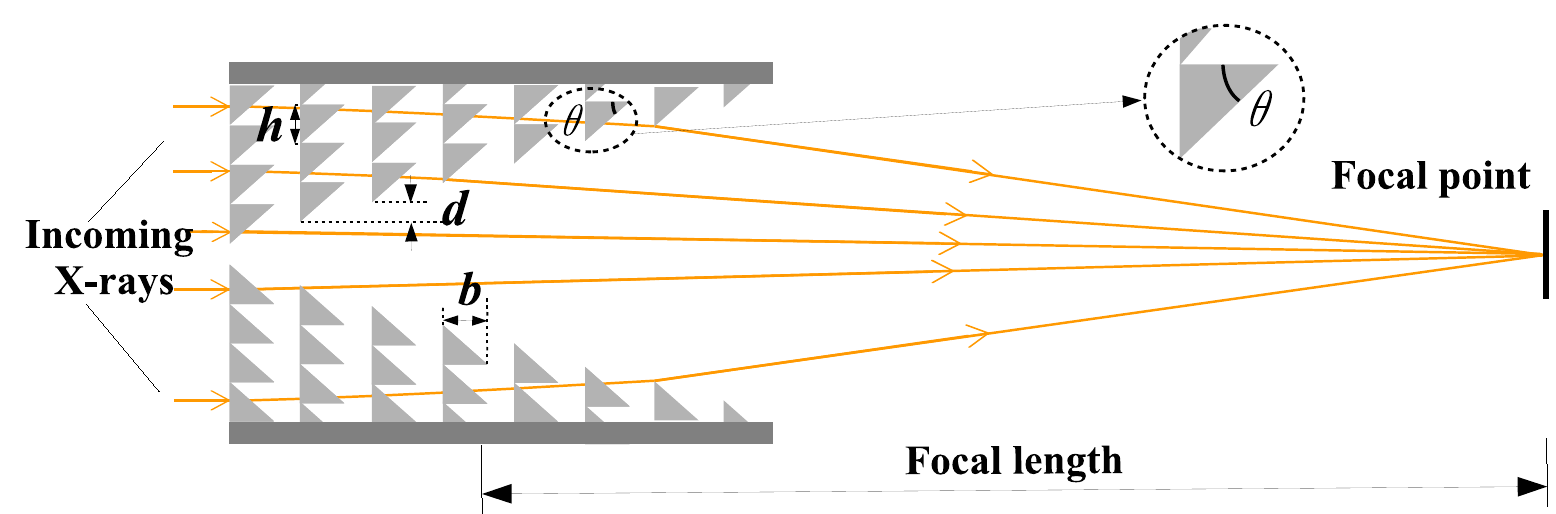}
	 \label{F:SPLscheme}}
    \bcaption{Illustration of the proposed Stacked Prism Lens.}{(\textbf{a}) The Stacked Prism Lens is a stack of discs with an increasing number of prisms toward the periphery, resulting in the desired focusing effect. (\textbf{b}) The cross section of one disc used to build Stacked Prism Lenses shows the embedded circular prismatic rings. (\textbf{c}) Principle of nested Wolter type I optics, consisting of  parabolic mirrors followed by hyperbolic mirrors to reflect and focus X-rays. The focal length is 10 m for the hard X-ray focusing telescope, NuSTAR, in the energy range 3 - 78.4 keV. (\textbf{d}) Principle of Stacked Prism Lens, consisting of identified prisms to refract and focus X-rays. $d$ is the displacement between two adjacent prism column, $h$ is the prism height and $b$ is the prism base. $\theta$ is the angle subtended by the prism height, i.e. $\tan^{-1} ( {h} /{b})$. The focal length can be less than 0.1 m.}
    \label{F:SPLlens}
\end{figure*}

Fig.~\ref{F:SPLlens} shows the schematics of the proposed Stacked Prism Lens. Each individual lens (Fig.~\ref{F:stacklens1}) is built by stacking discs embedded with a variable number of identical prismatic rings (Fig.~\ref{F:stacklens2}).

\floatsetup[figure]{style=plain,subcapbesideposition=top}

The lens can be described as a rotationally symmetric version of prism-array lenses\cite{cederstrom2005generalized} or Clessidra lenses\cite{jark2004focusing}. The basic scheme of the Stacked Prism Lens (Fig.~\ref{F:SPLscheme}) is obtained by removing chunks of material corresponding to a multiple of $2\pi$ phase shift in a multi-prism lens\cite{Cederstroem2000}, chosen to functionally approximate the shape of parabolic lens stepwise with linear segments\cite{nillius2011large}. Unlike Wolter type I optics (Fig.~\ref{F:wolterscheme}), these lenses focus incident radiation into a spot via both refraction (Fig.~\ref{F:SPLscheme}) and diffraction\cite{jark2008role}. The focal length $f$ of each Stacked Prism Lens is
\begin{equation}
f = {{d \tan{\theta}} \over {\delta}},
\end{equation}
where $d$ is the columnar displacement, $h$ is the prism height and $b$ is the prism base. $\theta$ is the angle subtended by the prism height, i.e. $\tan^{-1} ( {h} /{b})$. $\delta$ is the deviation from unity of the real part of the refractive index, $ n = 1-\delta + i\beta $. Similarly to the compound refractive lens, a shorter focal length can be obtained by stacking a series of Stacked Prism Lenses, i.e. $1/f = \sum 1/f_i$.

\subsection{{Fabrication of Stacked Prism Lenses.}}
\begin{figure}[ht]
 \centering
  \includegraphics[width = 8cm]{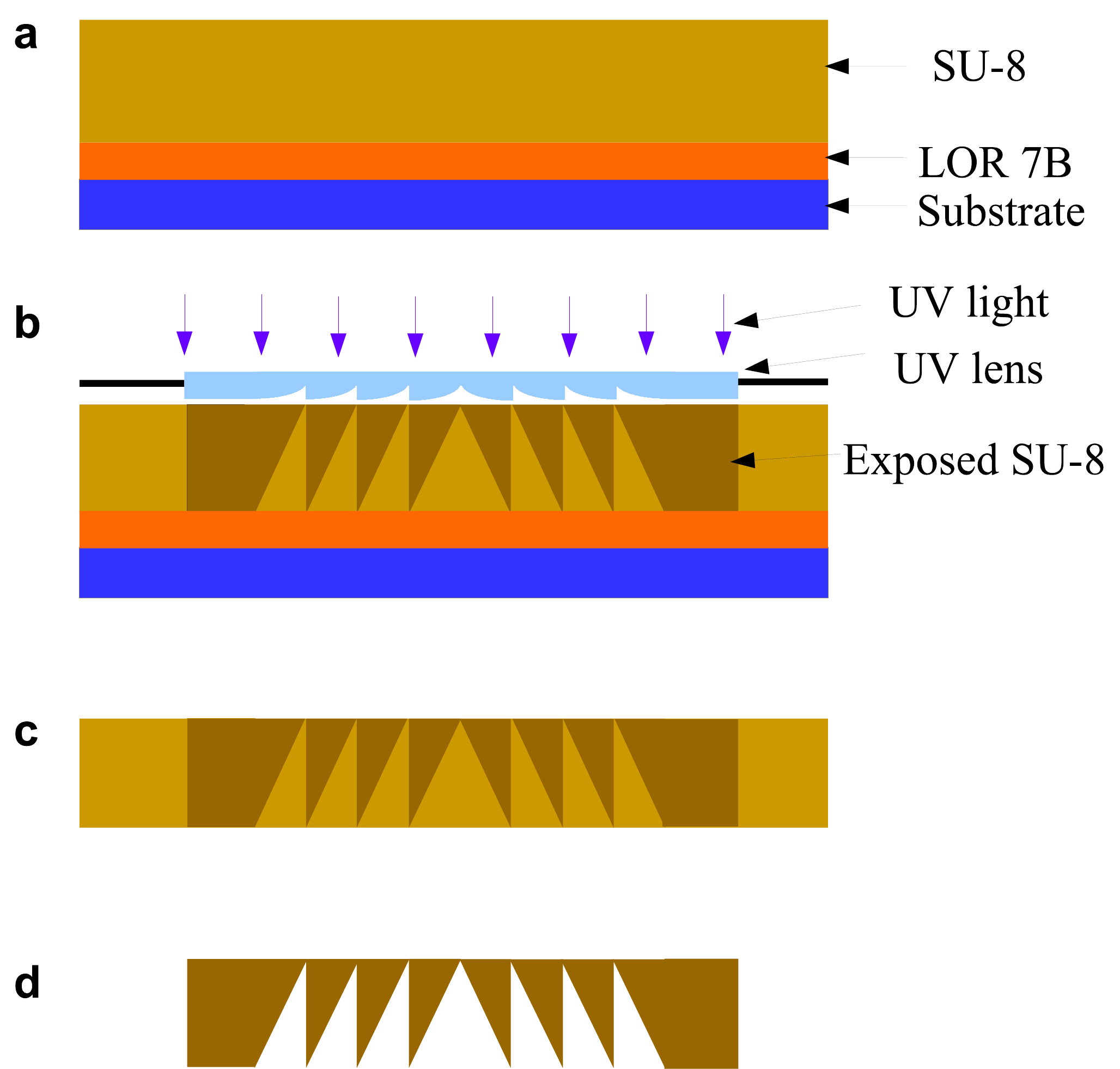}
  \bcaption{Main fabrication process steps of the discs for the Stacked Prism Lens. (not to scale)}{(\textbf{a}) A sacrificial layer LOR 7B and a photoresist SU-8 2100 layer are spin-coated on the silicon substrate. (\textbf{b}) The applied photoresist is exposed with focused UV lithography. (\textbf{c}) The exposed sample is pre-developed and released from the silicon substrate. (\textbf{d}) The released structure is developed in SU-8 developer.}
  \label{F:process}
\end{figure}

The main fabrication process steps for Stacked Prism Lenses are illustrated in Fig.~\ref{F:process}. 
The discs used to build up Stacked Prism Lenses are fabricated using focused UV lithography\cite{mi2016fabrication}, for which a UV lens is employed to shape the UV beam and pattern the photoresist, instead of the binary photomask used in conventional UV lithography. The UV lens is fabricated with grayscale e-beam lithography\cite{mi2014efficient}. The negative photoresist, SU-8 2100, is used as structural material for the discs. A v-groove with low surface roughness is utilized to align fabricated discs and assemble the Stacked Prism Lens. 

\begin{figure} 
  \centering 
   \sidesubfloat[]{ 
    \label{F:Lenseside} 
    \includegraphics[width = 9cm]{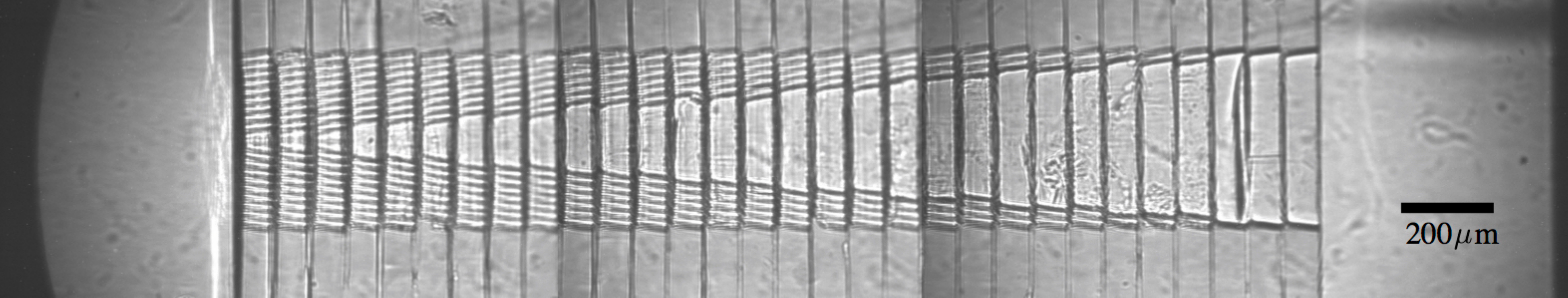}}\\
    \vspace{0.5cm}
  \sidesubfloat[]{ 
    \label{F:lens_recpart} 
    \includegraphics[width = 4cm]{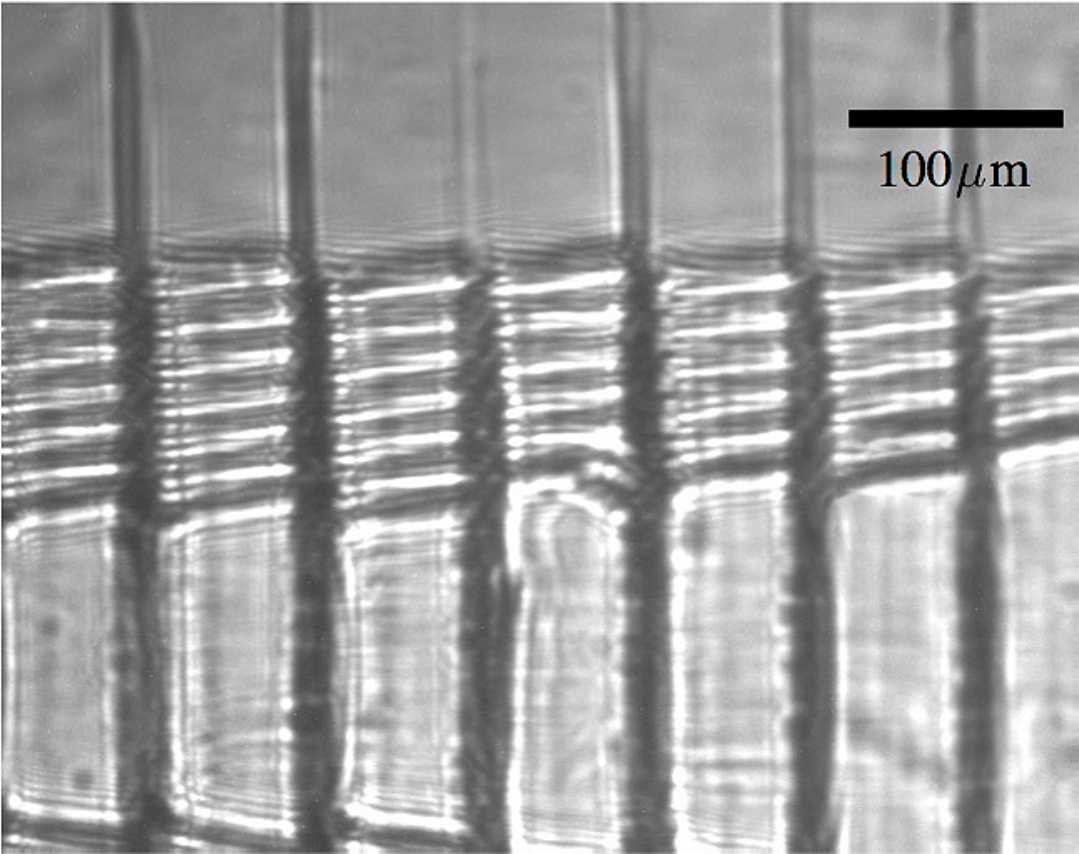}}
    \hspace{0.5cm}
  \sidesubfloat[]{ 
    \label{F:lens_light} 
    \includegraphics[width = 4.2cm]{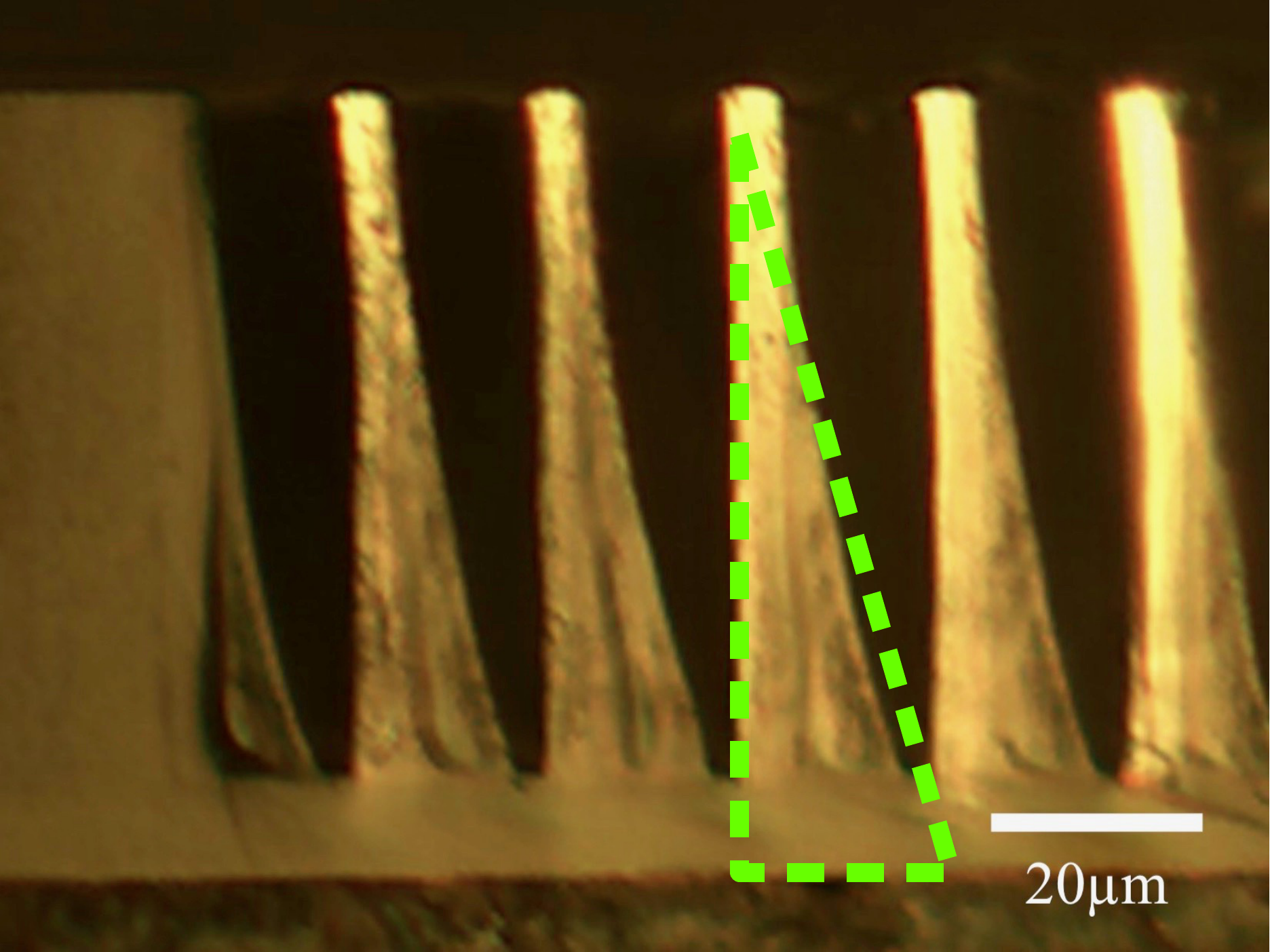}} 
    \bcaption{The fabricated discs and completed Stacked Prism Lens.}{(\textbf{a}) Transmission image of the fabricated lens, illuminated by 13.5 keV X-rays from the side and constructed by stitching three exposures from the detector together. (\textbf{b}) Detail of \textbf{a}, showing some structural collapse. (\textbf{c}) Cross-section image of partial fabricated disc under an optical microscope. The green dashed lines indicate the designed prism.} 
 \label{F:xraylens} 
\end{figure}

Fig.~\ref{F:Lenseside} and \ref{F:lens_recpart} 
show transmission images of a completed Stacked Prism Lens, which were taken by illuminating the lens from the side using an X-ray beam. The imaged lens comprises 30 discs with 1 mm diameter, embedded with different numbers of circular prismatic rings. The prism base $b = 84\ \si{\micro\meter}$; the prism height  $h = 20\ \si{\micro\meter}$; and the column displacement $d = 20\ \si{\micro\meter}$, in comparison to the design values with $b = 76\ \si{\micro\meter}$, $h = 20\ \si{\micro\meter}$ and $d = 20\ \si{\micro\meter}$. The geometric aperture of the lens is 420 $\si{\micro\meter}$. Fig.~\ref{F:lens_light} shows a cross-section image of a fabricated disc under an optical microscope. As can be expected in the first prototypes, there are some imperfections in the fabricated Stacked Prism Lens, such as structural collapse (Fig.~\ref{F:lens_recpart}) and obtuse prism tips (Fig.~\ref{F:lens_light}), which will degrade lens performance.
 
\subsection{Optical properties of Stacked Prism Lenses.}

\begin{figure*}[ht]
 \centering
   \sidesubfloat[]{
  	 \includegraphics[width = 10cm]{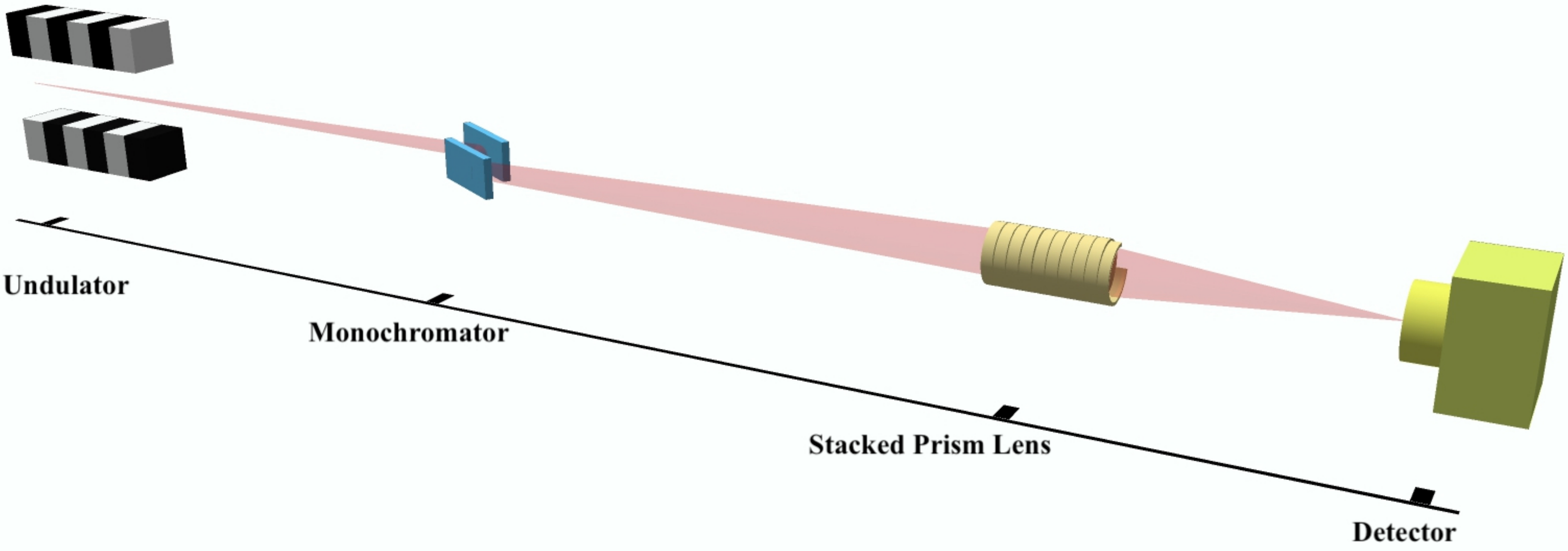}
  	 \label{F:setup}}
	 \vspace{0.1cm}
  \sidesubfloat[]{
  	 \includegraphics[width = 4.5cm]{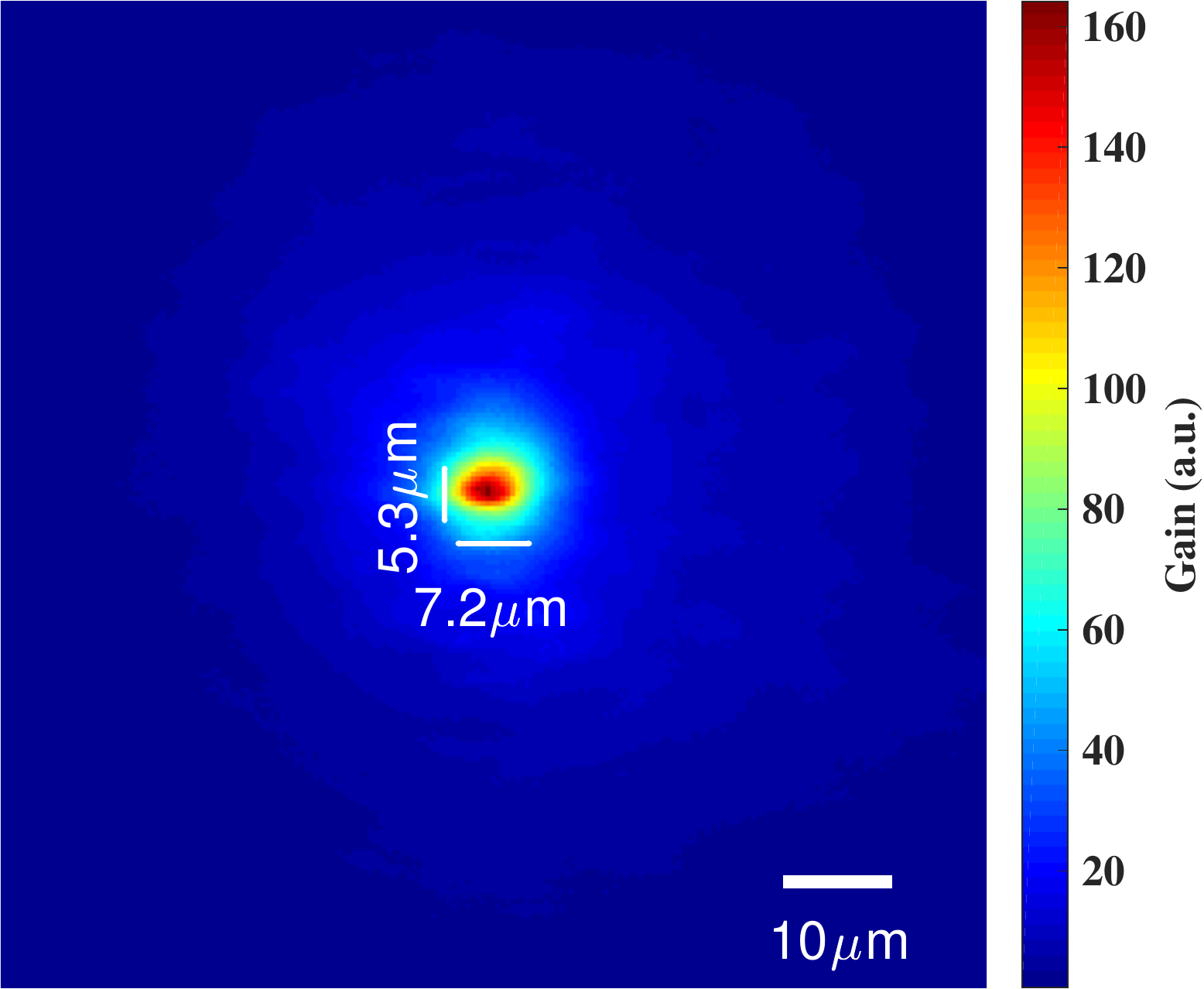}
  	 \label{F:gainFWHM}}
  \sidesubfloat[]{
  	 \includegraphics[width = 4.5cm]{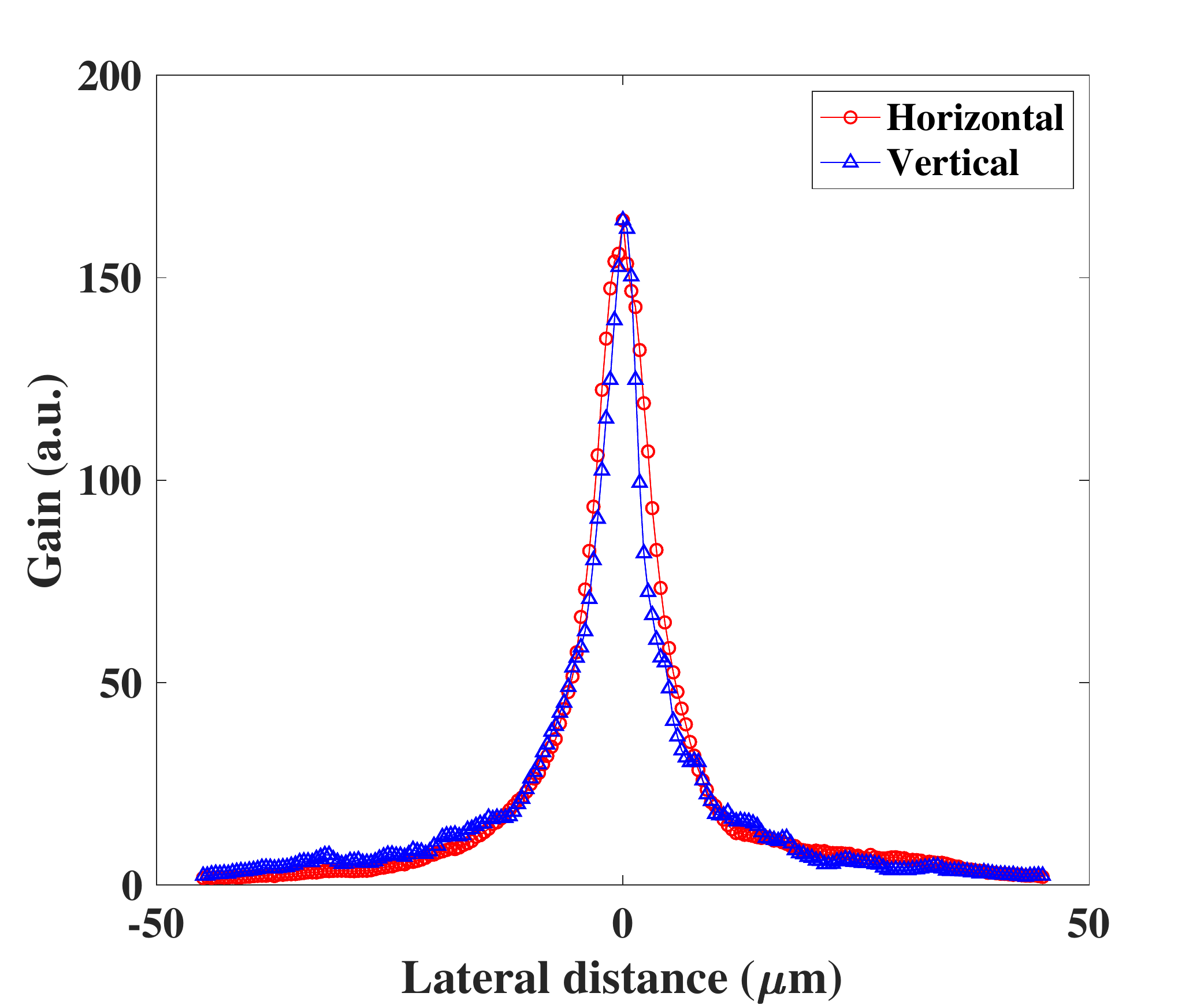}
  	 \label{F:FWHM}}
	 \\
  \sidesubfloat[]{
  	\includegraphics[width = 10cm]{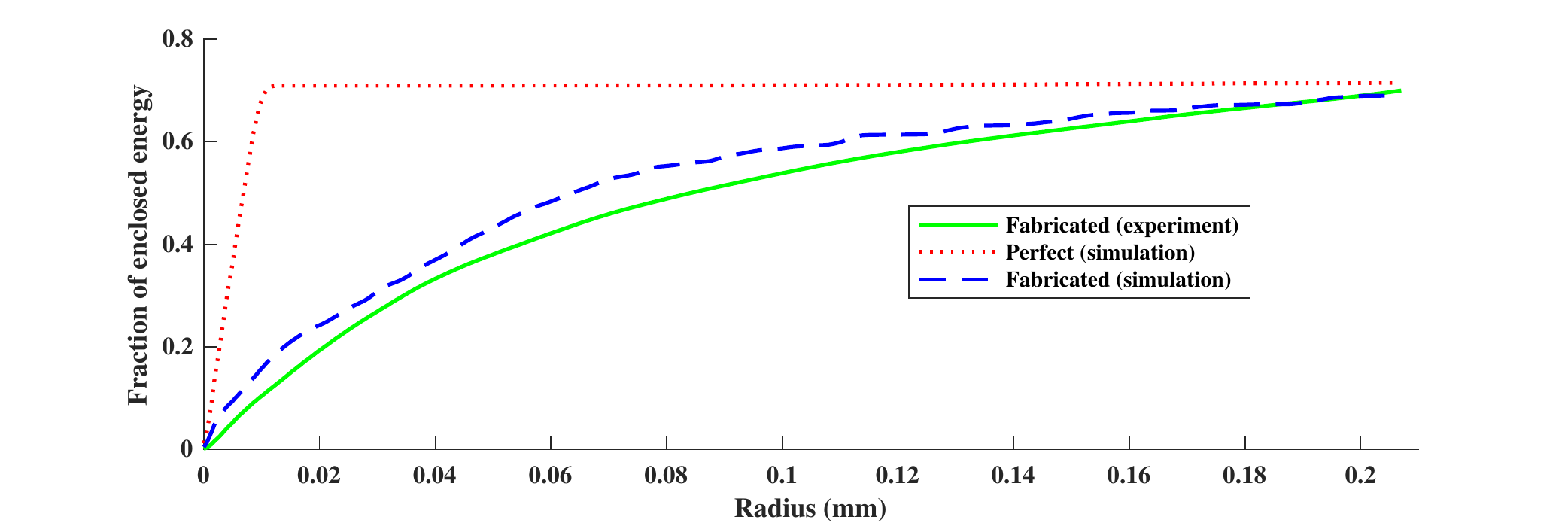}
	 \label{F:effvsR}}
    \bcaption{Experimental and simulated performance of the Stacked Prism Lens.}{(\textbf{a}) Experimental setup for Stacked Prism Lens optical properties measurement (not to scale). (\textbf{b}) Measured gain distribution for the fabricated lens working at 13.5 keV and lens-to-detector distance at 705mm.The focal spot is 7.2 \si{\micro\meter} (FWHM) horizontally and 5.3 \si{\micro\meter} (FWHM) vertically. (\textbf{c}) Vertical and horizontal slices through the focal spot. (\textbf{d}) The fraction of energy enclosed versus radius for different Stacked Prism Lenses. The solid green line represents the experimental result for the fabricated lens, the dashed blue line represents modelled fabricated lens, and the dotted red line represents the modelled ideal lens. The experimental error bars are smaller than the line width in the figure.}
\end{figure*}

The optical performance of the fabricated Stacked Prism Lens was characterized at beamline B16 of the Diamond Light Source (Oxford, UK), with the experiment setup outlined in Fig.~\ref{F:setup}. This beamline has a source with an effective spot size of 126 \si{\micro\meter} $\times$ 56 \si{\micro\meter} (FWHM). The fabricated Stacked Prism Lenses were illuminated with a monochromatic beam as the lens-to-beam rotation angle, titled angle and lens-to-detector separation, and the beam energy were scanned iteratively to find the configuration for the best performance. A PCO.4000 camera with a pixel size of 9 \si{\micro\meter}, a $20 \times$ objective and a point spread function (PSF) with an FWHM of 2.3 \si{\micro\meter} was used to record the intensity distribution. In addition, a ray-tracing model was used to simulate the optical performance of Stacked Prism Lenses with both ideal and manufactured prism profiles assuming the same configurations as in the experiment. The simulation was written specifically for this purpose and implemented in a combination of Matlab and C++. To compare with the measurements the simulated intensity is convolved with the detector's point-spread function.

The minimum spot size and maximum intensity gain were obtained at an energy of 13.5 keV (Supplementary Fig.~\ref{F:experiment}), for a lens-to-detector separation of 705 mm, compared to 717 mm for both simulation and numerical calculation results. The gain distribution in the focal plane (Fig.~\ref{F:gainFWHM}) is calculated by comparing intensity distributions with and without a lens. Just like the X-ray source, the measured focal spot is elliptical with FWHM of 7.2 \si{\micro\meter} and 5.3 \si{\micro\meter} in the horizontal and vertical direction, respectively, and the maximal intensity gain is around 164. The measured FWHM is around 50\% wider and the gain is less than one third of the simulated values. Deconvolving with X-ray spot size and the detector's PSF, the real spot size (FWHM) is estimated to be around 3 \si{\micro\meter}, which is far from the numerically calculated diffraction-limited resolution of 152 nm. These differences can be attributed to the deficiencies (such as prism profile, alignment, and so on) between the simulation model and the fabricated lens.

The focal properties for Stacked Prism Lens are analysed in more detail in Fig.~\ref{F:effvsR}, which shows the fraction of enclosed energy as a function of radius, obtained by integrating the intensity inside a certain position from the optical axis and comparing to the total illuminated flux. Besides resolution, efficiency is also an important property for hard X-ray optics. For the fabricated lens, the measured total efficiency is about 70\%, compared to 69\% and 71.5\% for the simulation results of the fabricated and ideal lens, respectively. This means that the effective aperture is about $84\%$ of the geometric aperture, 352 \si{\micro\meter}. (For an ideal beryllium compound refractive lens with the same focal length and geometric aperture, the total efficiency is only 15\%.) However, the energy inside the central peak (with a radius of around 10 \si{\micro\meter}) is about 10\%, which means only about 15\% of the transmitted flux is refracted properly into the central peak. 50\% of the transmitted flux is distributed over an area with a radius of around 43 \si{\micro\meter}, compared to around 35 \si{\micro\meter} and 5 \si{\micro\meter} for the simulation results of fabricated and ideal lenses, respectively.

The optical properties of the tested Stacked Prism Lens are slightly worse than the simulation result for the manufactured prism profile, but are not compatible with the simulation for an ideal prism profile. According to the simulations, the performance degradation is mainly due to profile aberration and alignment error, both arising from the equipment which is currently available in our lab. In the future, we will improve the fabrication processes by reducing the UV wavelength for lithography to avoid overexposure of tips (Supplementary Fig.~\ref{F:uvlens}) and using an accurate alignment machine when assembling. We expect that this will allow fabricated lenses to reach the theoretical performance values.

\subsection{Hard X-ray telescope concept based on Stacked Prism Lens array.}

\begin{figure}[h]
  \centering
  \sidesubfloat[]{
  	 \includegraphics[width = 9cm]{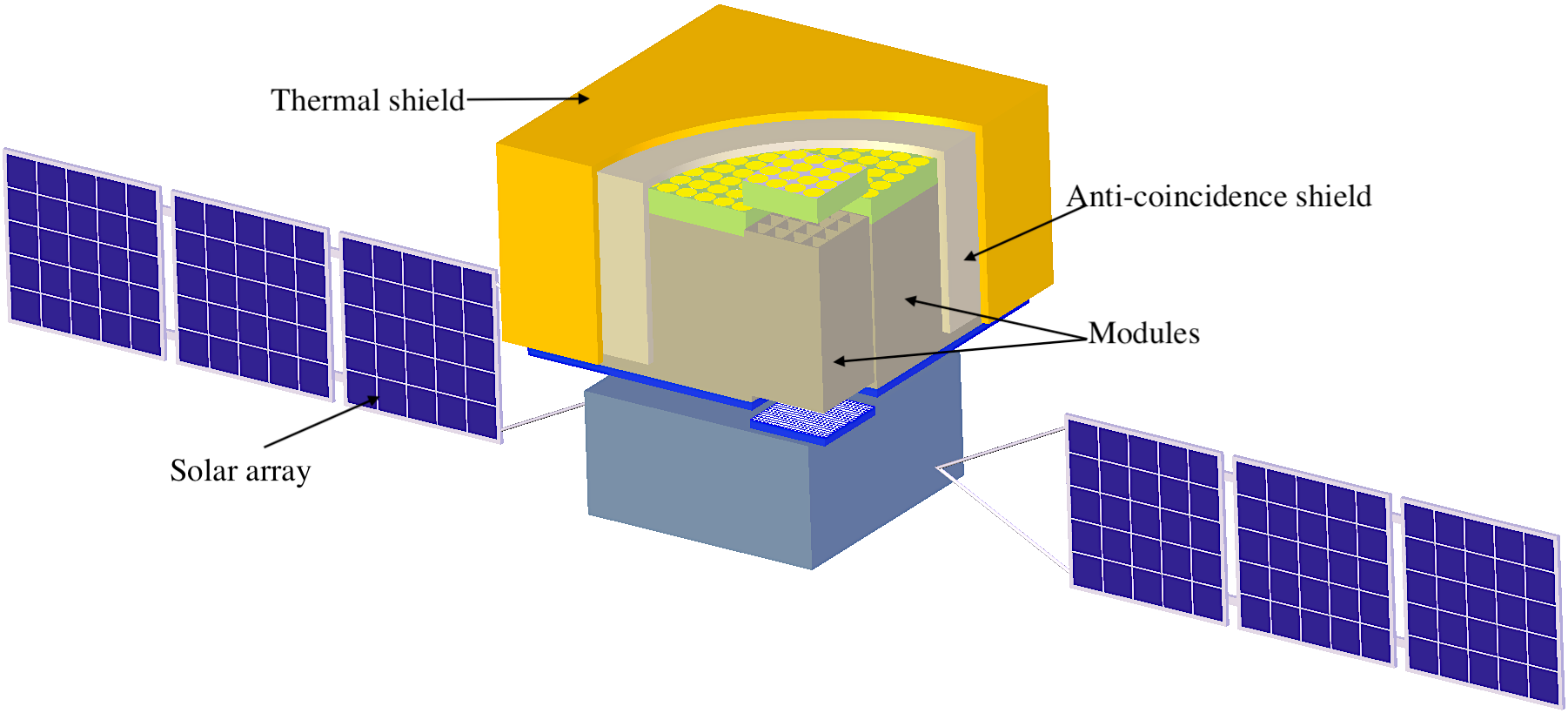}
  	 \label{F:telescope}}
 \sidesubfloat[]{
  	 \includegraphics[width = 2.7cm]{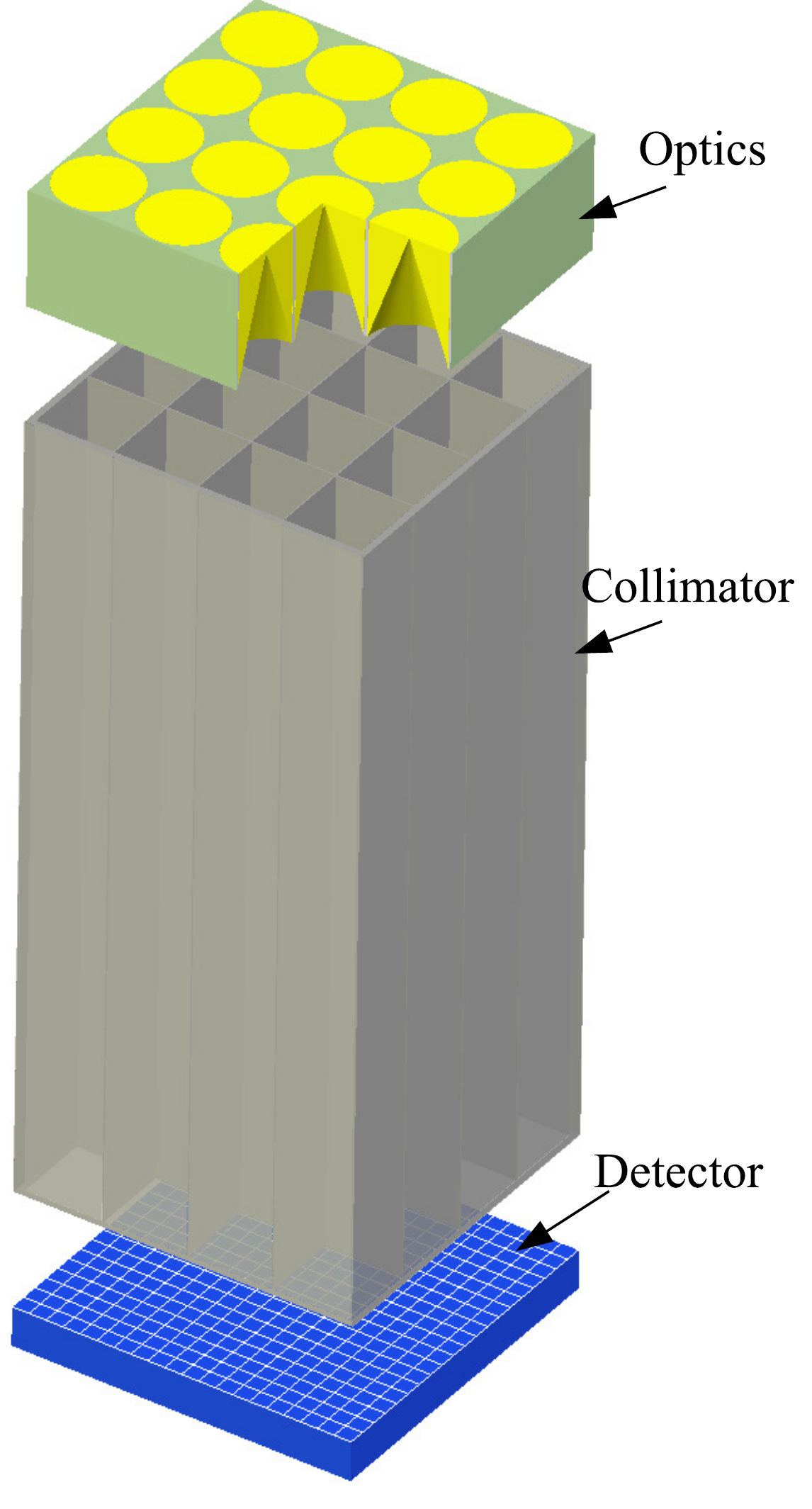}
  	 \label{F:module}}
	 \hspace{5cm}
  \sidesubfloat[]{
  	\includegraphics[width = 5cm]{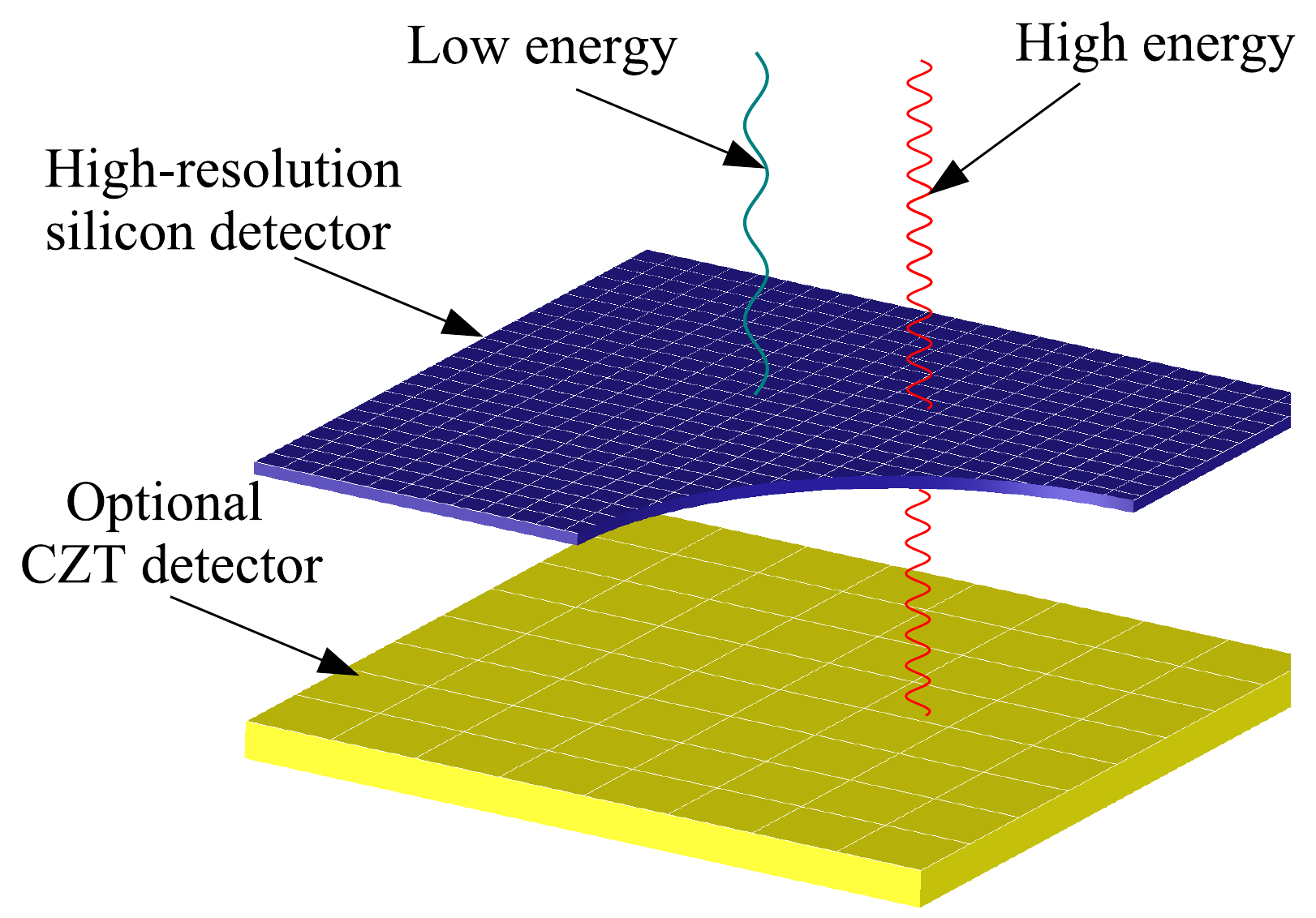}
	 \label{F:detector}}
	 \hspace{0.5cm}
   \sidesubfloat[]{
  	\includegraphics[scale=0.5 ]{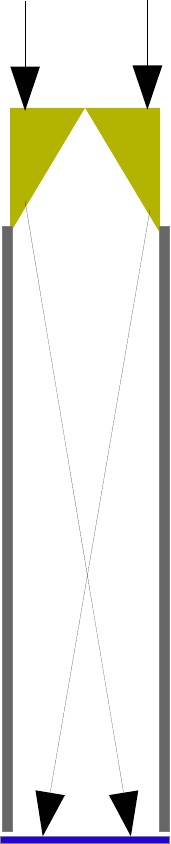}
	 \label{F:low}}
	 \hspace{0.1cm}
    \sidesubfloat[]{
  	\includegraphics[scale=0.5]{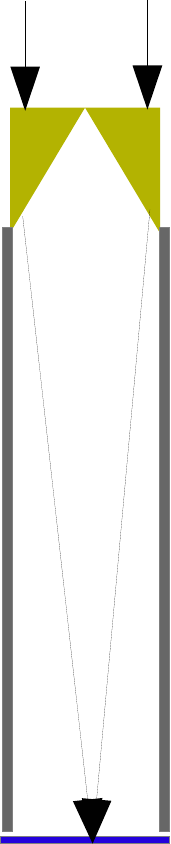}
	 \label{F:design}}
	 \hspace{0.1cm}
    \sidesubfloat[]{
  	\includegraphics[scale=0.5]{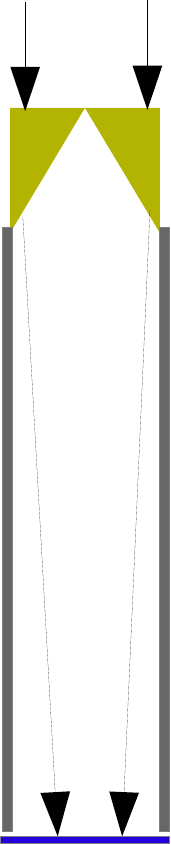}
	 \label{F:high}}
	 \hspace{0.1cm}
    \sidesubfloat[]{
  	\includegraphics[scale=0.5]{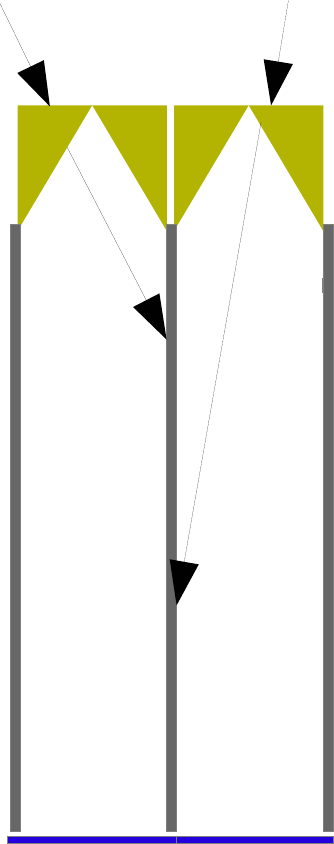}
	 \label{F:CT}}
    \bcaption{Implementation of a hard X-ray focusing telescope based on the Stacked Prims Lens array.}{(\textbf{a}) Hard X-ray focusing telescope with modules in a parallel arrangement. All the modules can be covered by a thermal shield and segmented anti-coincidence shield. (\textbf{b}) A single module in \textbf{a}, consisting of an optics plane, a detector plane, and collimators. Each module is cuboid shaped with a square base of side length 0.04 m. The height reflects the focal length which lies in the range 0.1-0.5 m. (\textbf{c}) Proposed double-layer detector for X-rays of different energies. (\textbf{d}) Ray path for X-rays with energy lower than optimum energy. (\textbf{e}) X-rays of optimum energy will be focused onto the detector plane with high resolution. (\textbf{f}) Ray path for X-rays with energy higher than optimum energy. (\textbf{g}) Rays with larger incident angles will be blocked and absorbed by the collimator.}
    \label{F:HXT}
\end{figure}

The basic concept for a hard X-ray focusing telescope based on the Stacked Prism Lens array is shown in Fig.~\ref{F:HXT}, 
for which the instrumental part is built by an array of modules in a parallel arrangement (Fig.~\ref{F:telescope}). Each module (Fig.~\ref{F:module}) has its own optics plane and detector plane, supported by a multi-grid array collimator. Each grid element works as a miniature hard X-ray focusing telescope, consisting of a corresponding Stacked Prism Lens, a detector unit and a collimator aperture.

As is illustrated above, the optics plane consists of a Stacked Prism Lens array, manufactured by focused UV lithography using SU-8. The spatial resolution in half power diameter (HPD, defined as the diameter of the region that encompasses 50\% of the focused flux) is around 4 \si{\micro m} at the optimum energy (Supplementary Fig.~\ref{F:XRTlens}). The collimator comprises Tungsten or Molybdenum perforated with thousands of precisely aligned holes. The wall stops scattered stray X-rays and prevents crosstalk between adjacent grids (Fig.~\ref{F:CT}). We propose a 1 mm thick active pixel Silicon sensor with 2 \si{\micro m} spatial resolution\cite{fossum1997cmos} and a partly event driven read-out, for a target power consumption of less than 1 \si{\micro W/mm^2}. An alternative sensor design is a backlit CCD\cite{soltau2000fabrication}. We do not expect the sensor to limit the performance of the telescope. A second sensor layer (Fig.~\ref{F:detector}) of 3 mm CZT would extend the upper energy limit to 200 keV, with coarser spatial resolution. 

Compared with grazing incidence optics, a telescope based on a Stacked Prism Lens array has many advantages (Table~\ref{T:para}). 
\begin{table}[htb!]
\centering
\captionsetup[table]{position=above}
{\caption{The performance of the hard X-ray focusing telescopes based on Wolter optics and the Stacked Prism Lens array. Stacked Prism Lens assemblies are formed into modules which allows a scaleable telescope to be designed. Assuming that the modules are 4 cm $\times$ 4 cm, the potential effective area is $\sim$$N$$\cdot$7.2 cm$^2$ and $N$$\cdot$8.5 cm$^2$ for SPLs with focal length 0.109 m and 0.54 m, respectively. $N$ is a scale factor, and its value depends on the size of the launcher payload fairing. The full potential of the SPL telescope will likely be reached through the development of a series of telescopes with successively increasing area. In order to explore the full potential of our approach, we consider a telescope which occupies half of the Falcon Heavy fairing (with an internal usable width of 4.6 meters for 6.7 meters of its total 11 meters internal length, tapering to 1.45 meters wide at the top of its usable height). In this scenario, the total number of modules for the 0.109 m (0.54 m) SPL is 180000 (83000), resulting in an effective collecting area of $\sim$ 130 m$^2$ ($\sim$ 70 m$^2$) once the modules are unfolded in orbit.} \label{T:para}}
{
\linespread{1.5}\selectfont
\renewcommand{\arraystretch}{1}
\resizebox{.9\textwidth}{!}{
\begin{tabular}{ l c c c c }
\toprule
 {Parameters} & NuSTAR & HXT(ASTRO-H) & XRT based on SPLs & XRT based on SPLs \\
\midrule
 Optics & Wolter-I & Wolter-I &\tabincell{c}{SPL array\\Diameter: 0.31 mm} & \tabincell{c}{SPL array\\Diameter: 1 mm} \\  
 \hline
 Resolution HPD & 60$''$ & $\sim$ 110$''$ & \tabincell{l}{11$''$@13.5 keV\\389$''$@80 keV} & \tabincell{l}{1.6$''$@13.5 keV\\261$''$@80 keV} \\  
 \hline
 Resolution FWHM & 18$''$ & {$<$0.4$''$}& 0.1$''$@13.5 keV & 0.04$''$@13.5 keV \\ 
 \hline
 FoV&12.5$'$&9$'$&9.8$'$ &6.4$'$\\
 \hline
 \tabincell{c}{Maximum Effective \\area (cm$^2$)}&900&1 200&$\sim$$N\cdot$7.2 @13.5 keV&$\sim$$N$$\cdot$8.5 @13.5 keV\\
 \hline
 Energy range (keV) &3-78.4& 5-80 &5-100&5-100\\
 \hline
 Focal length (m)&10.15&12&0.109&0.54\\
 \hline
 \(\displaystyle \frac{\textrm{Effective area}}{\textrm{Geometric area}} \)& \tabincell{l}{0.39@10 keV\\0.11@30 keV\\0.04@60 keV}&\tabincell{c}{0.38@10 keV\\0.11@30 keV\\0.03@60 keV}&\tabincell{c}{0.27@10 keV\\0.45@13.5 keV\\0.82@60 keV}&\tabincell{c}{0.32@10 keV\\0.53@13.5 keV\\0.85@60 keV}\\
 \hline
 Detector&\tabincell{c}{CZT}&{CZT}& Si/CZT&Si/CZT\\
 \hline
 \tabincell{c}{Minimum mass per unit \\effective area (kg/m$^2$)}&\tabincell{c}{$\sim$940}&$>$940 &$\sim$223 &$\sim$187\\
 \bottomrule
\end{tabular}
}
}
\end{table}
One advantage is higher angular resolution. The simulation suggests that for Stacked Prism Lenses with an ideal prism geometry, angular resolution better than 5$''$  (HPD) and 0.1$''$ (FWHM) can be achieved, a significant improvement compared to NuSTAR  (Supplementary Fig.~\ref{F: ASvsenergy}) and HXT(ASTRO-H). The PSF distorts as the off-axis angle increases, but the HPD remains approximately constant inside the FoV defined by the collimator (Supplementary Fig.~\ref{F:offaxis}). The use of linear prisms in the SPL results in the HPD significantly exceeding the FWHM (see discussion in Methods). The Stacked Prism Lens approach allows a modular hard X-ray focusing telescope to be designed. A major advantage of this approach is that the short focal length (e.g. $\sim$ 0.5 m, compared to $>$10 m for NuSTAR) means that a compact and lightweight ($<$200 kg/m$^2$) telescope is possible. There are many practical issues to resolve before realising the full scaleable potential of this approach. One such issue is the alignment and calibration of individual telescope assemblies within a module. We foresee that automated precision micro-fabrication techniques can be exploited here. When the full potential of the modular design is exploited, the resulting effective area and grasp (product of on-axis effective collecting area and effective FoV) stands to be significantly larger than offered by current missions (Supplementary Fig.~\ref{F: Evsenergy} and Fig.~\ref{F: grasp}).

As a variant of the parabolic refractive lens, a Stacked Prism Lens is intrinsically chromatic since the focal length is proportional to $\lambda^{-2}$ (Supplementary Fig.~\ref{F:focallength}), which means that high resolution is only achieved for X-rays with limited bandwidth around an optimum energy (Fig.~\ref{F:design}). For X-rays with lower and higher energies (Fig.~\ref{F:low} and~\ref{F:high}), the angular resolution is degraded (Supplementary Fig.~\ref{F: ASvsenergy}). Thus, it appears very difficult to match the performance of a Wolter type XRT across a broad X-ray spectrum. This problem can be partially addressed by combining Stacked Prism Lenses designed for different optimum energies or by combining a diffractive phase Fresnel lens and a Stacking Prism Lens to correct the chromatic aberration \cite{skinner2004design,wang2003achromatic}.

\ssection{Discussion}

In this paper, we have demonstrated the feasibility and characterization of the Stacked Prism Lens for hard X-rays and investigate a practical application. By approximating the parabolic lens but partially removing the absorption materials, we can construct novel X-ray optics with point focusing and high resolution, high efficiency and large effective aperture size. The manufactured Stacked Prism Lenses comprise SU-8 discs embedded with variable numbers of prismatic rings. These discs were patterned using focused UV lithography, a technique which is simple, inexpensive, and suitable for mass production.

Through experiment, we show that the manufactured Stacked Prism Lens is able to match their expected performance on resolution, efficiency, and effective aperture. Although there were fabrication imperfections, we were still able to attain a focal spot of 5 \si{\micro\meter} $\times$ 7 \si{\micro\meter} (FWHM),  total transmission of 70\%, and a large effective aperture of 352 \si{\micro\meter}, for a geometric aperture of 420 \si{\micro\meter} at an optimum energy of 13.5 keV. Simulation studies suggest that significantly improved results can be achieved for ideal prisms with improved alignment between the prismatic ring elements. The fabrication process needs to be improved to achieve this.

A hard X-ray focusing telescope concept is introduced using an array of Stacked Prism Lenses and a two-layer detector arrangement. Compared with today's state-of-the-art hard X-ray telescopes, it is more compact (due to the short focal length) and lightweight (with high efficiency and light polymer lenses leading to small mass per effective collecting area) with the modular approach having the potential to provide a significant increase both in effective area and grasp. Depending on the application the angular resolution can be improved around an optimum design energy over current Wolter optics, while sacrificing resolution towards higher energies. Despite drawbacks in terms of bandwidth, there is excellent scientific potential for a hard X-ray telescope based on Stacked Prism Lenses in hard X-ray astronomy as well as numerous applications in other areas, such as condenser lenses in X-ray microscopes\cite{marschall2014x}, optics for HyperSPECT system\cite{tibbelin2012simulation}, and so on.

\ssection{Acknowledgements}
We acknowledge the Diamond Light Source for provision of synchrotron radiation facilities and express our thanks to Oliver Fox and Kawal Sawhney for assistance in applying and using beamline B16. We thank Cheng Xu and Staffan Karlsson for taking part in the experiment at the Diamond Light Source, Christer Svensson for the detector design and power calculation and Le Mi for X-ray telescope concept drawing. W.M and M.D. acknowledge funding from Stiftelsen Olle Engkvist Byggmästare. M.P. acknowledges funding received from the Swedish Research Council (grant number 2016-04929).
\ssection{Author contributions} 
W.M. performed the design, fabrication, test experiments and simulation of the Stacked Prism Lens, analyzed the experiment data and space application, and prepared the paper. P.N. contributed to the design of Stacked Prism Lens and the development of the simulation program, performed the X-ray experiments, and discussed the experiment results. M.P. discussed and developed the space application of Stacked Prism Lenses and assisted with the writing of the manuscript. M.D.  managed the project, proposed applications for space and other areas, and together with the co-authors coordinated the writing of the manuscript. 
\ssection{Author information}
Reprints and permissions information is available at www.nature.com/reprints. The authors declare that they have no competing financial interests. Correspondence and requests for materials should be addressed to Wujun Mi (email: wujunmi@mi.physics.kth.se).

\ssection{Methods}

\subsection{Lens investigated.} 
Stacked Prism Lenses use prisms or curved prisms to focus X-rays. Each prism has a height, h, and a base, $b = m \lambda / \delta(\lambda)$, where $m$ is a positive integer to ensure the phase continuity, $\lambda$ is the wavelength, and $\delta(\lambda)$ is the refractive index decrement from unity for an X-ray of wavelength $\lambda$.
The beam deflection angle when passing a single prism is\cite{cederstrom2002multi}
\begin{equation}
    \Delta = \delta(\lambda)/ \tan{\theta}
  \end{equation}
 The rays, hitting the lens at a distance $r$ from optical axis, will be deflected through an angle
 \begin{equation}
    \Delta(r) = {\big [{{r} \over {d}} \big ]}  {\delta(\lambda) \over \tan{\theta}}
  \end{equation}
where the square brackets denote rounding to the nearest integer. The rays near the edges pass through more prisms than the rays in the centre, thereby creating a focusing effect. However, the rays hitting the lens within the same columnar displacement, will be refracted with the same angle, i.e. perfect focusing is not possible with these linear prisms. A promising solution to this issue is to use prisms with a curved profile\cite{jark2004focusing}. For the Stacked Prism Lens, $d$ is a free parameter and is not limited by the prism height. Smaller $d$ means a closer approximation to an ideal Fresnel pattern and thus smaller focal spot, but requires more accurate alignment and complicated fabrication. Similarly with the compound refractive lens, shorter focal length can be obtained by stacking several identical Stacked Prism Lenses, which results in 
   \begin{equation}
    F = {{d \tan{\theta}} \over {N\delta(\lambda)}}
  \end{equation}
Here, $N$ is the number of Stacked Prism Lenses. Since $\delta \propto \lambda^2$, the focal length $F \propto \lambda^{-2}$ or $F \propto E^2$ . The projected material thickness of Stacked Prism Lenses at a distance to the optical axis $r$ can be expressed by
  \begin{equation}
    X(r) \approx {{m\lambda \tan{\theta} } \over {2\delta(\lambda)^2 F}} r = Kr
  \end{equation}
This shows that the thickness grows linearly with the distance to the optical axis. Taking geometric scattering\cite{nillius2011large} into consideration results in the transmission function
  \begin{equation}
    T(r) = e^{-K\mu r-{{r^2} \over {2\delta(\lambda) F^2} }}
  \end{equation}
where $\mu$ is the attenuation coefficient in the lens material, including Compton scattering and photo-electric absorption. The effective aperture, defined as the diameter of aperture that would transmit an equal amount of power throughout the whole geometric aperture, is an important parameter for X-ray optics. For Stacked Prism Lenses, it can be expressed by\cite{nillius2012geometric}
  \begin{equation}
    A_{eff} =2\sqrt{2\int_0^{A/2}{T(r)rdr}} 
    \label{E:A}
  \end{equation}
And the maximum aperture is computed as 
  \begin{equation}
    A_\infty =\lim_{A\to\infty}A_{eff}
    \label{E:AM}
  \end{equation} 
  The corresponding diffraction-limited resolution\cite{lengeler1999imaging}  is 
   \begin{equation}
    R = 0.75{{\lambda F} \over{A_{eff}}}
    \label{E:R}
  \end{equation} 

\subsection{Lithographic process for the UV lens.} 
The fabrication method for the UV lens is grayscale e-beam lithography\cite{mi2014efficient,mi2016fabrication}. The e-beam resist SML 1000 was used as structural material. The substrate was ITO-coated glass, which is almost transparent for near ultraviolet light. Before applying the SML 1000, a 40 nm-thick gold layer was deposited underneath the UV lens to form solid disks. A 1350 \si{\micro m}-thick layer of SML 1000 was then spin coated, baked at 180 \si{^{\circ}C} for 3 minutes and then patterned with a 25 keV Raith 150 e-beam lithography system. A proximity-effect-correction method based on multivariate adaptive regression splines (MARS) was employed to calculate the variable dose required for the desired UV lens profile. The patterned resist was developed in methyl isobutyl ketone (MIBK) for 20 seconds, rinsed in isopropyl alcohol (IPA) for 30 seconds, and then dried with flowing nitrogen. 

\subsection{Lithographic process for discs.} 
The stacked discs were fabricated by focused UV lithography with SU-8 2100 photoresist. The fabrication process started with a Si wafer of 550 \si{\micro m} thickness. After pretreating, LOR 7B and SU-8 2100 were spin-coated on the wafer with a thickness of 1 \si{\micro m} and 84 \si{\micro m}, respectively. The lift-off resist LOR 7B was utilized as a sacrificial layer to release the SU-8 structures from the Si wafer. After relaxation, the sample was baked on a levelled hotplate at 65 \si{^{\circ}C} for 3 hours. After cooling to room temperature, it was exposed with a Karl Suss MJB3 mask aligner equipped with an i-line filter in hard contact with the UV lens for 50 seconds. After post-exposure baking for 2 hours at 65 \si{^{\circ}C}, the sample was then pre-developed in SU-8 developer PGMEA (Propylene Glycol Monomethyl Ether Acetate) for 10 seconds with the help of ultrasonic vibration. The SU-8 structure was released from the Si wafer using Microposit\textsuperscript{{TM}} MF-319 developer and subsequently developed again in PGMEA at -20 \si{^{\circ}C} for 1 hour, rinsed in IPA for 30 seconds and then dried at room temperature.

\subsection{Lens simulation.} 
The software used to simulate the lens is custom-built, based on ray tracing and designed specifically for Stacked Prism Lenses. The regular structure of the Stacked Prism Lens is exploited to speed up the calculations, by storing the prisms in a regular grid structure, where each grid cell contains a small subset of the prisms. The prisms' shapes are represented as rotationally symmetric surfaces, i.e. each side of the prism forms a conical surface in 3D. Intersections between the photon path and the surfaces are computed analytically, thus avoiding the need to approximate curved surfaces with piece-wise linear polygon meshes, and instead obtaining full precision at this step. At each surface intersection, the refraction and reflection are computed according to Snell's law. Thereby, the full path of a photon can be traced through the lens and the optical system. The transmitted photons are gathered at a detector layer, where they are binned according to their position on the detector and the amount of attenuation. The attenuation of a photon path is computed according to the distance traveled in each material. Both photo-electric absorption and Compton scattering are considered. The Compton scattered photons are assumed be absorbed by the high-aspect-ratio collimator. The software is implemented in a combination of Matlab and C++ and the framework has been used to compare measurements with simulations in several previous works\cite{nillius2012geometric, nillius2011large}.
\subsection{Data Availability}
The data that support the plots within this paper and other findings of this study are available from the corresponding author upon reasonable request.

\ssection{References}
\bibliographystyle{naturemag}
\bibliography{mybiborg}

\newpage

\ssection{{Supplementary Figures}}  
\setcounter{figure}{0}   

\captionsetup[figure]{name={Supplementary Figure},labelsep={bar}}

\begin{figure*}[h]
   \sidesubfloat[]{
  	 \includegraphics[width = 12cm]{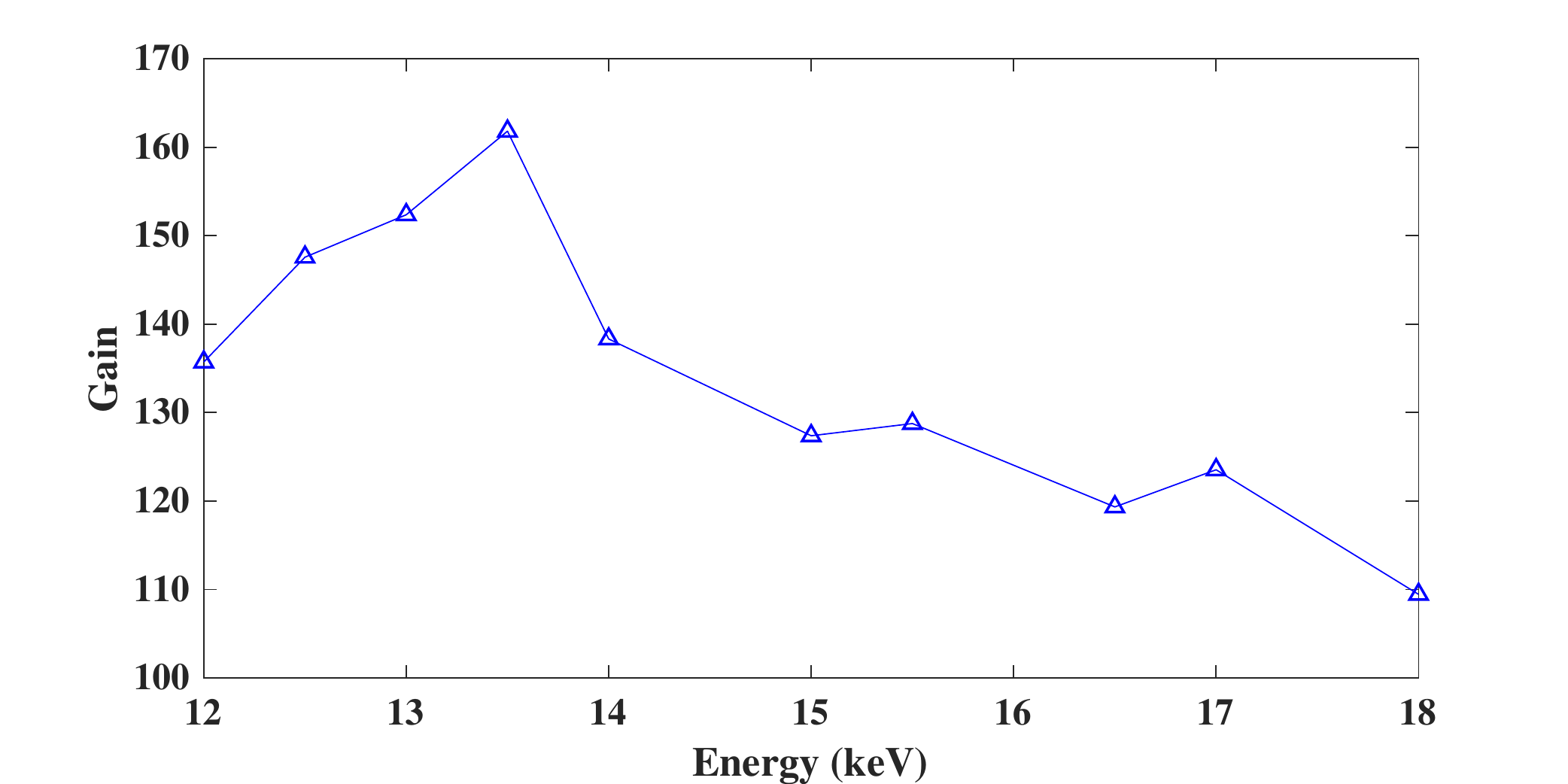}
  	 \label{F:gainvsE}}
	 \hspace{4cm}
  \sidesubfloat[]{
  	 \includegraphics[width = 12cm]{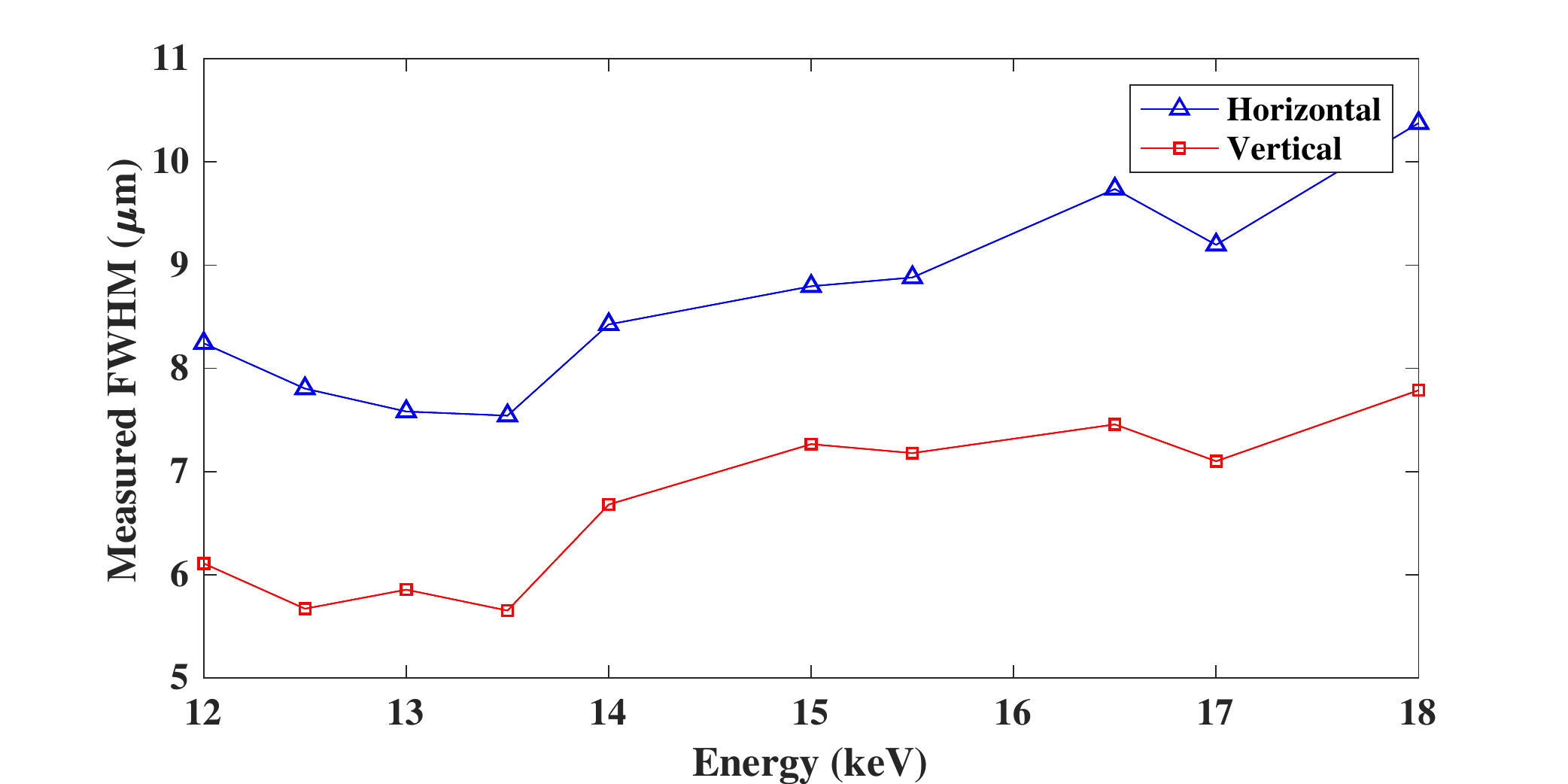}
  	 \label{F:FWHMvsE}}
    \bcaption{Energy dependence of the fabricated Stacked Prism Lens performance.}{(\textbf{a}) The measured intensity gain. The maximum value of 164 is found at 13.5 keV. (\textbf{b}) The measured FWHM. A minimum spot-size of 7.2 \si{\micro\meter} (vertical)  $\times$ 5.3 \si{\micro\meter}  (horizontal) is also at 13.5 keV.}
    \label{F:experiment}
\end{figure*}

\begin{figure*}[h]
 \centering
   \sidesubfloat[]{
  	 \includegraphics[width = 11cm]{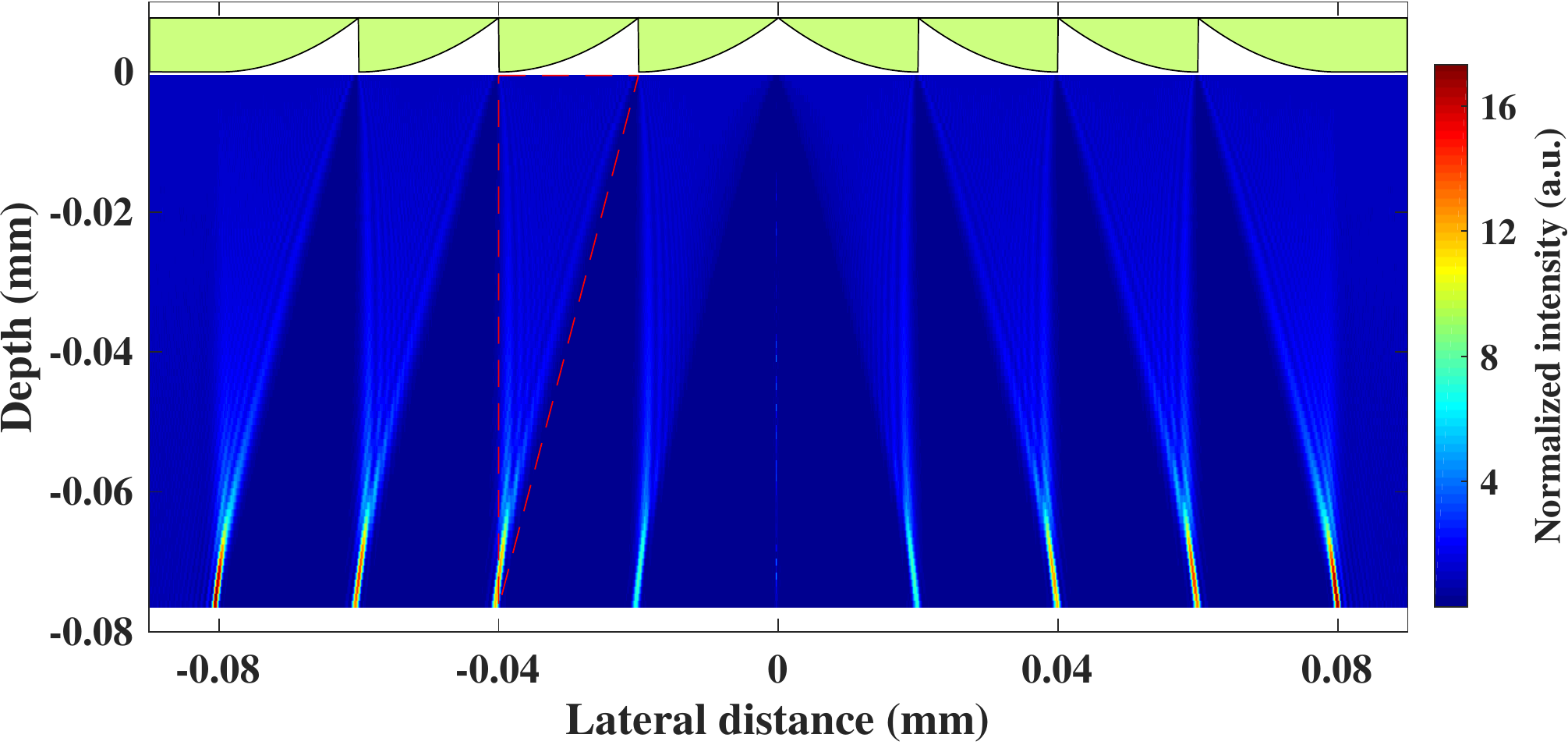}
  	 \label{F:gainvsE}}
	 \hspace{4cm}
  \sidesubfloat[]{
  	 \includegraphics[width = 11cm]{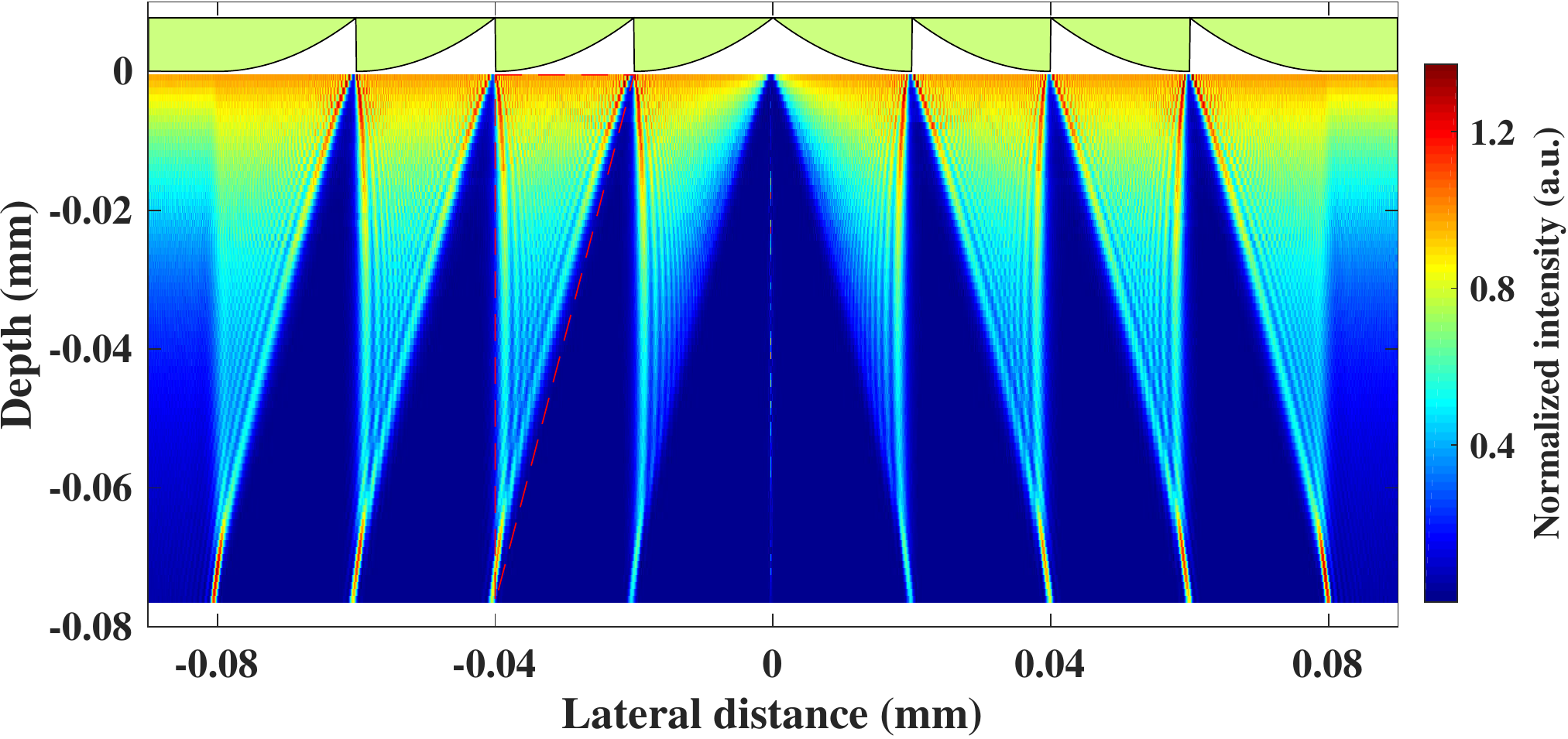}
  	 \label{F:FWHMvsE}}
    \bcaption{The exposure distribution in SU-8 for different UV wavelengths.}{(\textbf{a}) Exposure distribution for UV wavelength 365 nm. The exposure in the tips is more than ten times higher than that in the bases. (\textbf{b}) Exposure distribution for wavelength 345 nm. The exposure is more uniform throughout the whole thickness.}
    \label{F:uvlens}
\end{figure*}

\begin{figure*}[h]
 \centering
   \sidesubfloat[]{
  	 \includegraphics[height = 5cm]{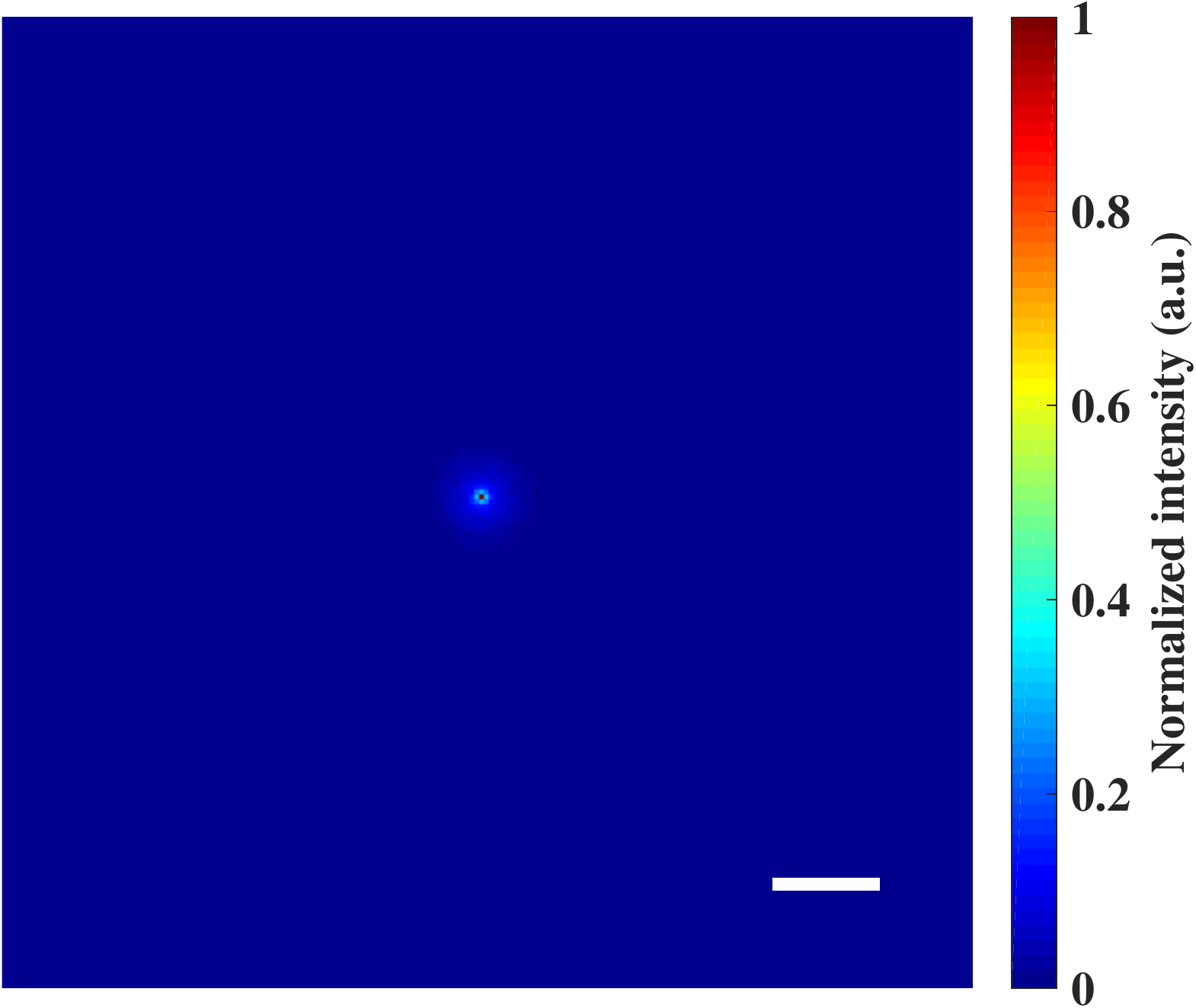}
  	 \label{F:XRT5i}}
  \sidesubfloat[]{
  	 \includegraphics[height = 5cm]{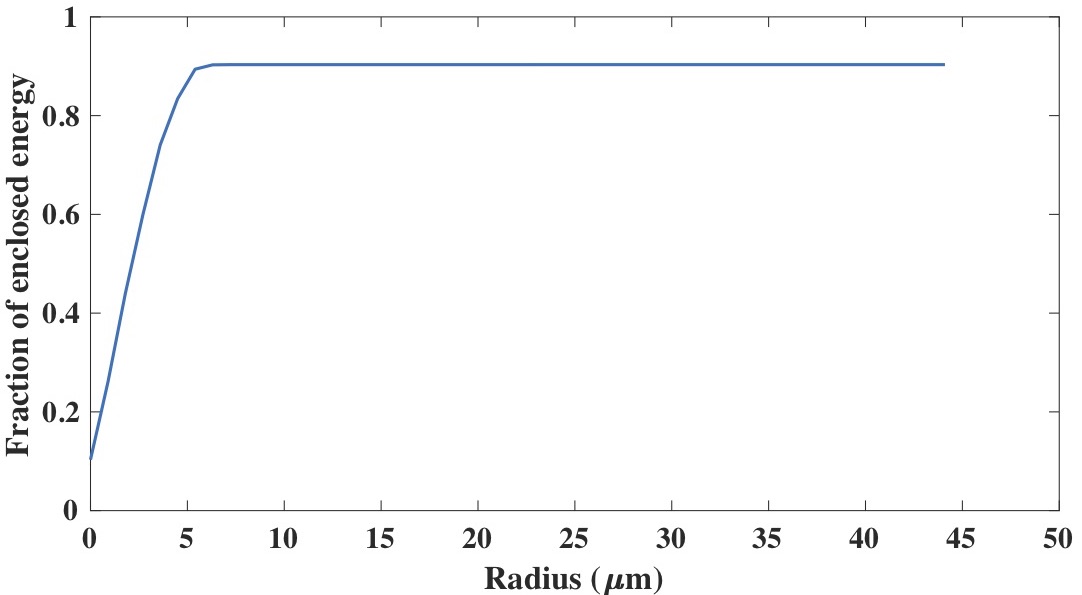}
  	 \label{F:XRT5h}}
\hspace{0.01cm}	 
  \sidesubfloat[]{
  	 \includegraphics[height = 5cm]{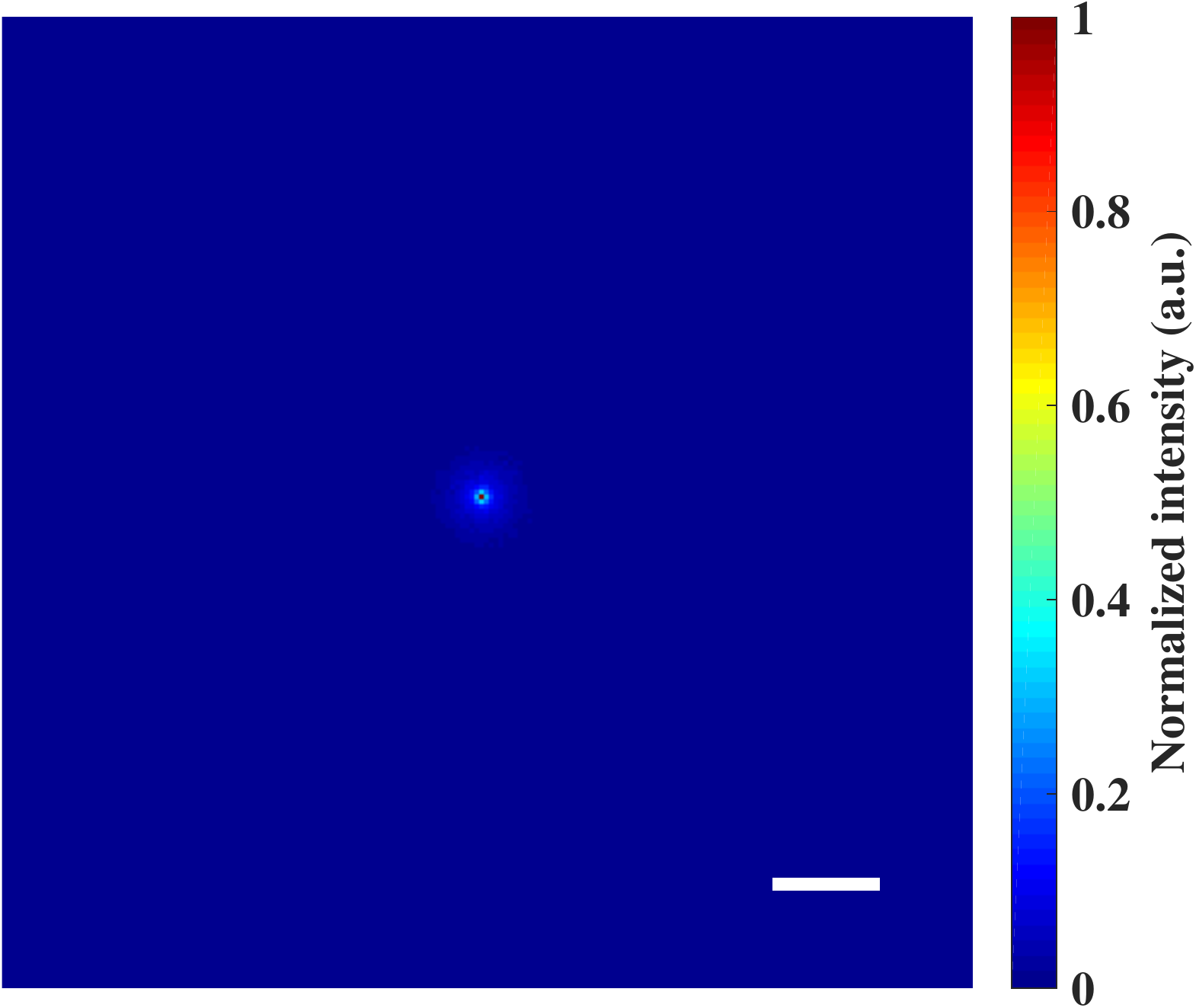}
  	 \label{F:XRT1i}}
  \sidesubfloat[]{
  	 \includegraphics[height = 5cm]{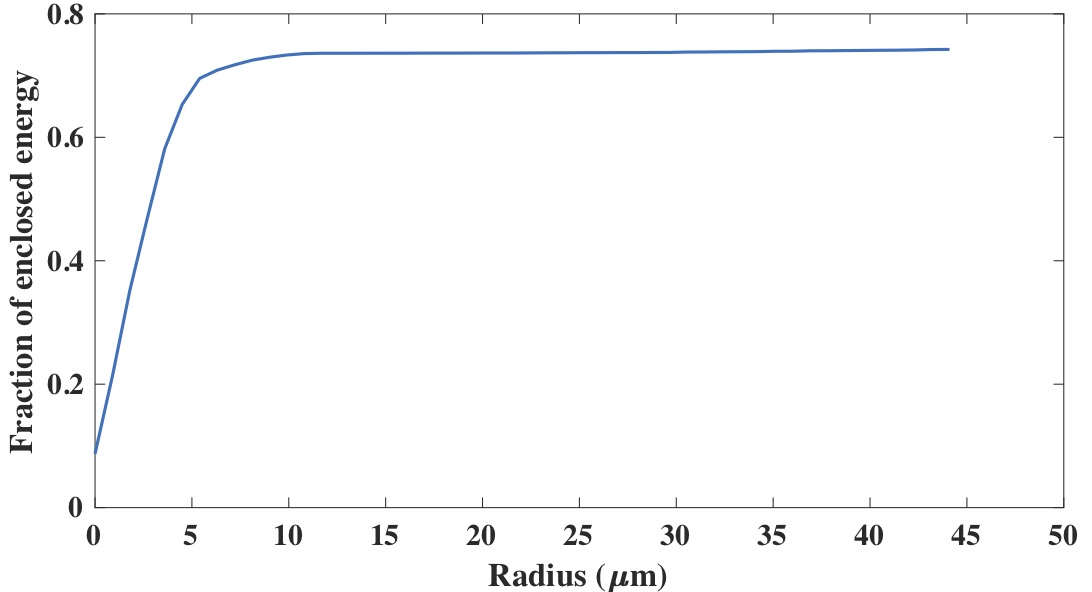}
  	 \label{F:XRT1h}}
    \bcaption{Optical performance of Stacked Prism Lenses suitable for an X-ray telescope optimized for 13.5 keV.} {(\textbf{a}) The focal plane intensity distribution for a Stacked Prism Lens with focal length 540 mm and a 1 mm aperture. The lens is built by stacking 2 identical Stacked Prism Lenses. Each lens has prism height $h = 10\ \si{\micro\meter}$, prism base $b = 62\ \si{\micro\meter}$ and columnar displacement $d = 10\ \si{\micro\meter}$. The scale bar represents 10 \si{\micro\meter}. (\textbf{b}) The fraction of energy enclosed for different radii in \textbf{a}. The HPD is 4.2 \si{\micro\meter}, corresponding to an angular resolution of 1.6$''$. (\textbf{c}) The focal plane intensity distribution for a Stacked Prism Lens with focal length 109 mm and an aperture of 0.31 mm. The lens is built by stacking 10 identical Stacked Prism Lenses. Each lens is designed with prism height $h = 10\ \si{\micro\meter}$, prism base $b = 62\ \si{\micro\meter}$ and columnar displacement $d = 10\ \si{\micro\meter}$. (\textbf{d}) The fraction of energy enclosed for different radii in \textbf{c}. The HPD is 5.8 \si{\micro\meter} corresponding to an angular resolution of 11$''$.}
     \label{F:XRTlens}
\end{figure*}

 \begin{figure*}[h]
 \centering 
 \includegraphics[width = 12cm]{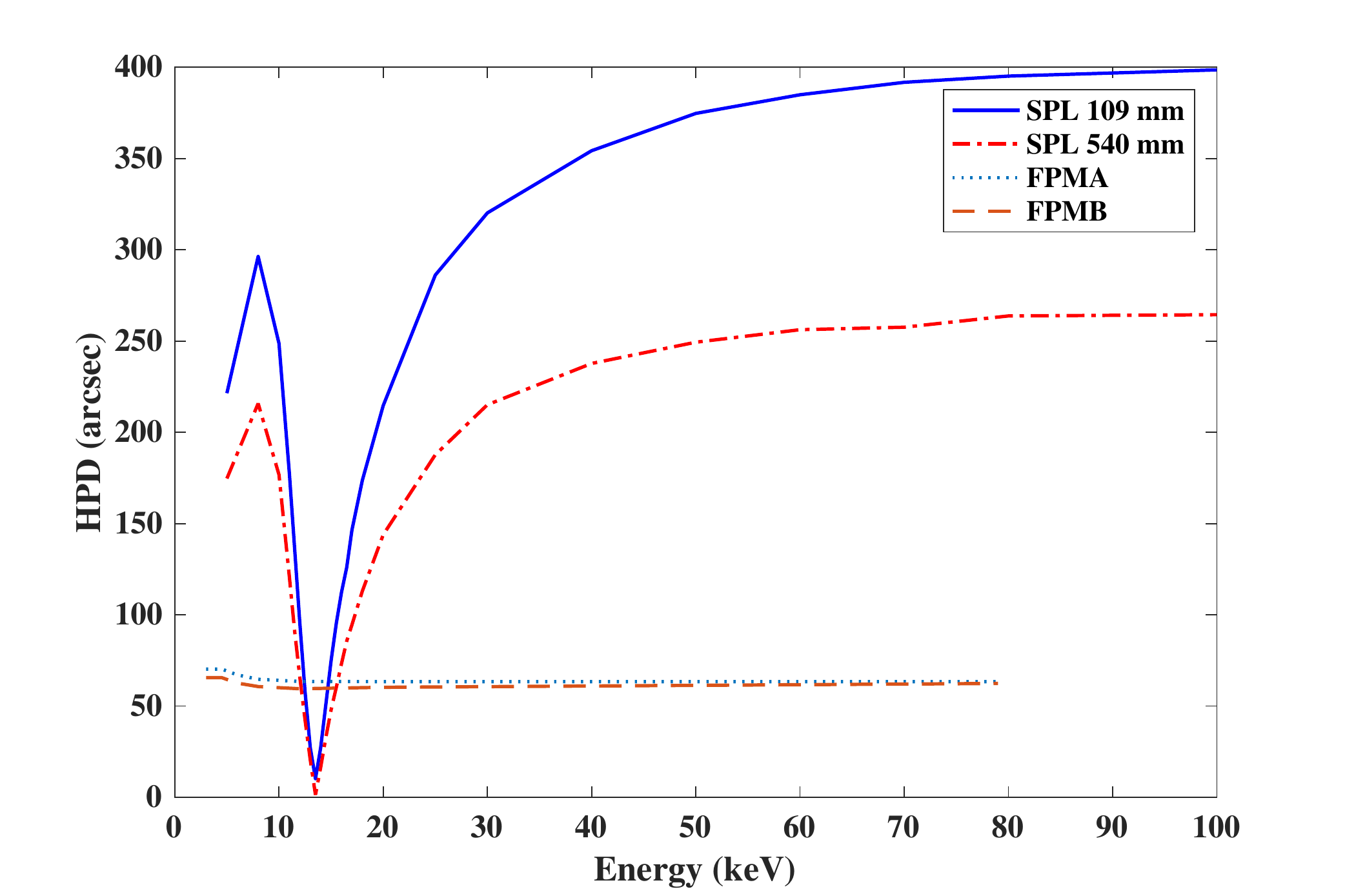}
    \bcaption{Simulated on-axis angular resolution (HPD) as a function of energy for Stacked Prism Lenses with different focal lengths and apertures, compared to the NuSTAR mission.}{The blue line denotes a Stacked Prism Lens with focal length 109 mm and an aperture size of 0.31 mm, while the red line denotes a lens with focal length 0.54 m and an aperture size of 1 mm. The Stacked Prism Lenses have high angular resolution around the optimum energy of 13.5 keV. Stacked Prism Lenses with different optimum energies can be combined to achieve broad-band imaging. The angular resolution for the two NuSTAR hard X-ray focusing telescopes (FPMA and FPMB) is around 60$''$ for energy range 3 - 78.4 keV.}
    \label{F: ASvsenergy}
\end{figure*}

\begin{figure*}[h]
 \centering
   \sidesubfloat[]{
  	 \includegraphics[width = 15cm]{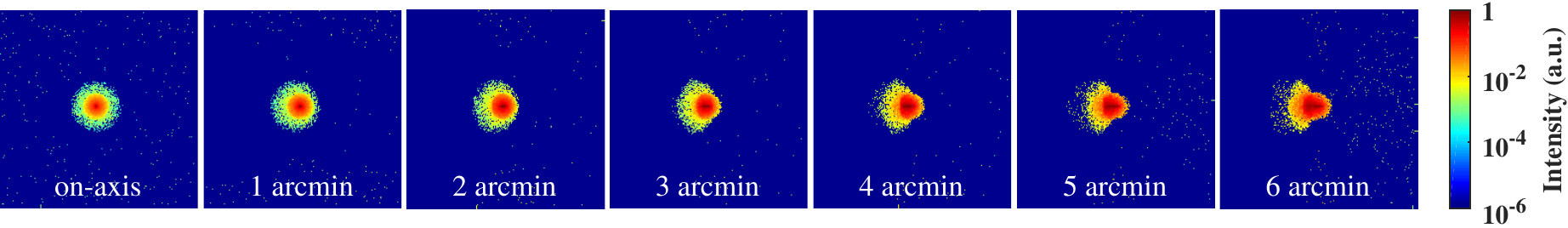}
  	 \label{F:po109}} 
	 
  \sidesubfloat[]{
  	 \includegraphics[width = 15cm]{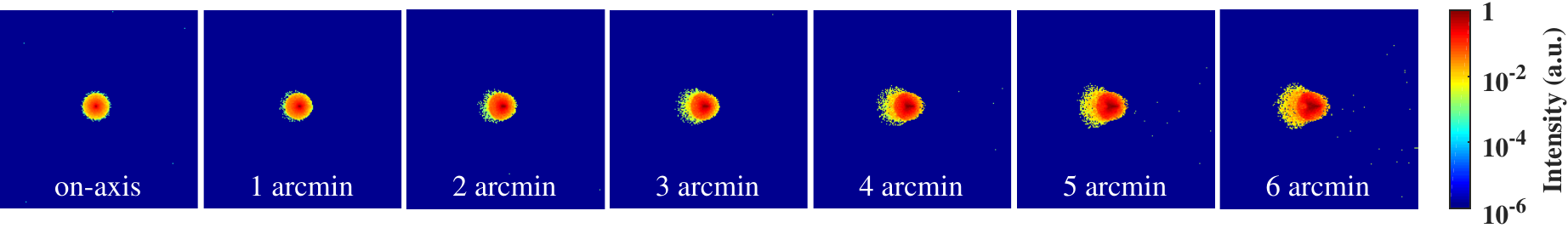}
  	 \label{F:po540}} 	 
	 	 
  \sidesubfloat[]{
  	 \includegraphics[width = 7.15cm]{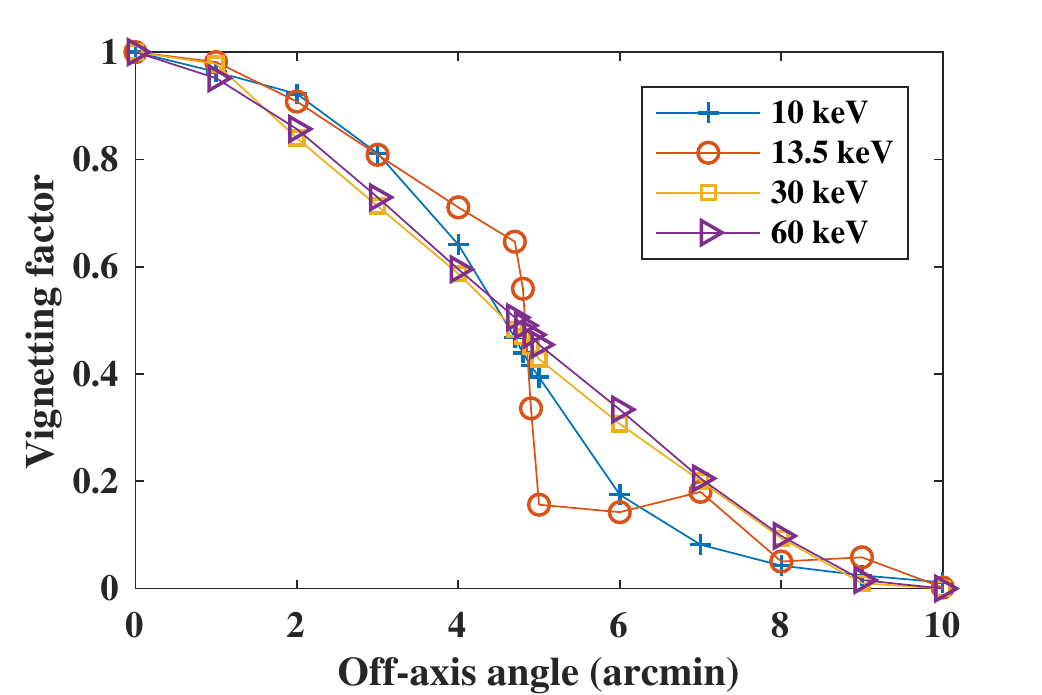}
  	 \label{F:vig109}}
 \sidesubfloat[]{
  	 \includegraphics[width = 7.15cm]{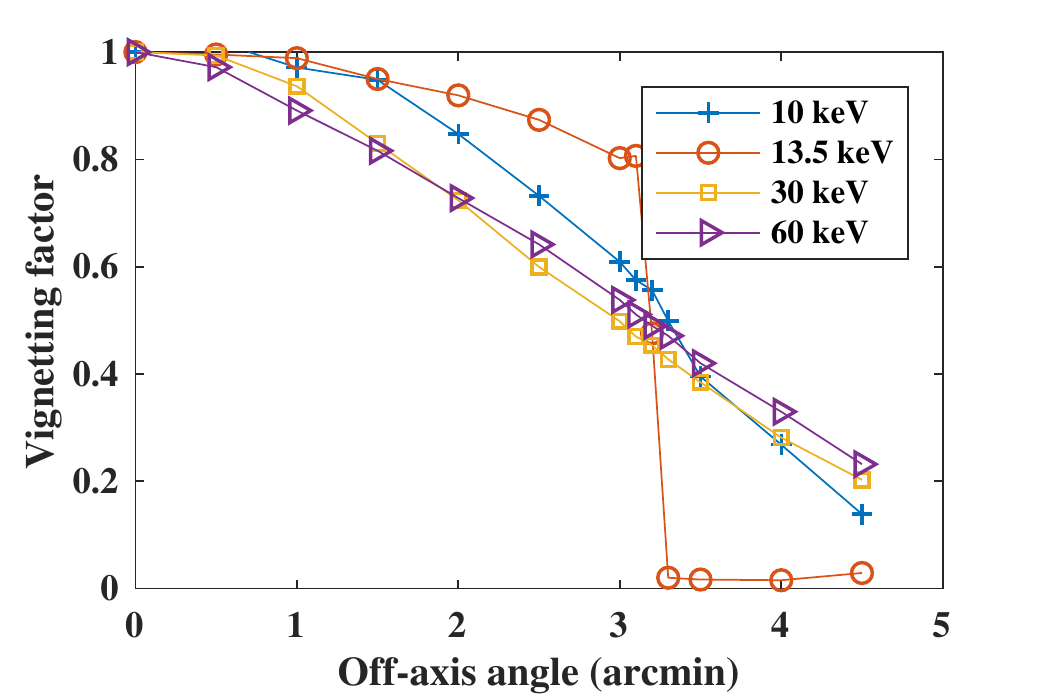}
  	 \label{F:vig540}} 
	 
 \sidesubfloat[]{
  	 \includegraphics[width = 11cm]{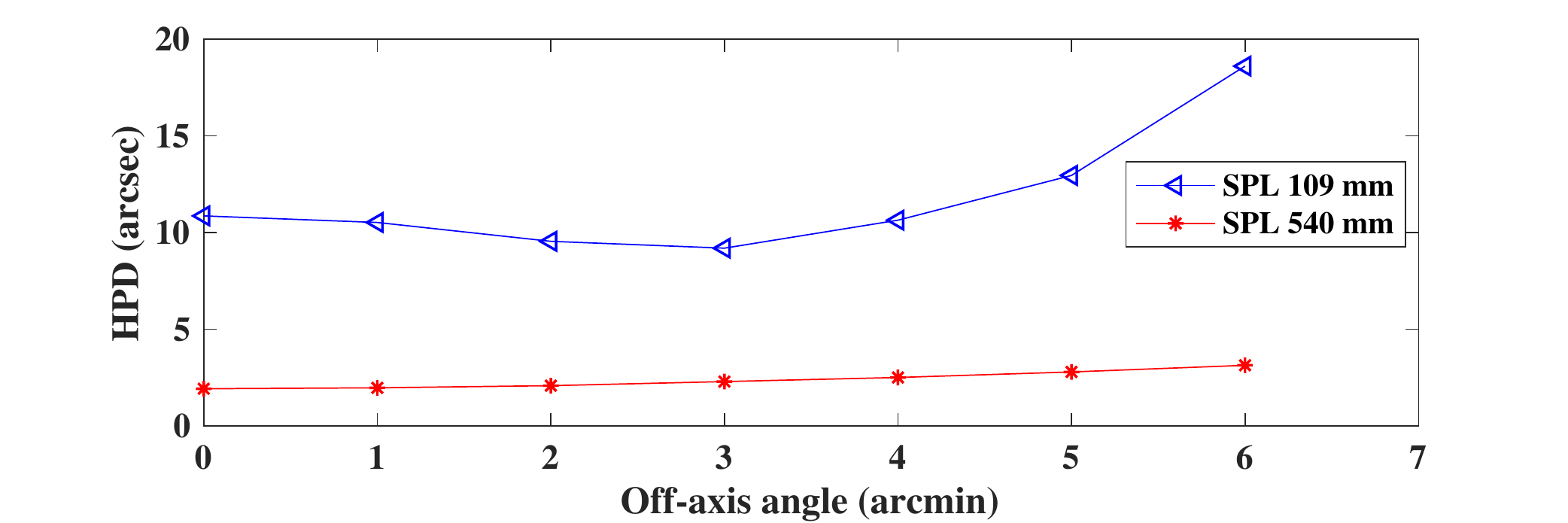}
  	 \label{F:ho}}	
    \bcaption{Simulated off-axis performance of Stacked Prism Lenses.} {(\textbf{a}) Normalized PSF images (logarithmic scale color) at different off-axis angles relative to the optical axis for SPL 109 mm. The size of each image is 90 \si{\micro\meter} $\times$ 90 \si{\micro\meter}. (\textbf{b}) Normalized PSF images (logarithmic scale) for SPL 540 mm. (\textbf{c}) The vignetting factor (defined as the ratio of the off-axis effective area and on-axis effective area for different X-ray energies) of a telescope based on SPL 109 mm for different X-ray energies. The effective FoVs (defined as twice the off-axis angle that has 50\% of the on-axis effective area) are 9.1$'$, 9.6$'$, 9.2$'$ and 9.5$'$ at 10 keV, 13.5 keV, 30 keV and 60 keV, respectively. (\textbf{d}) The vignetting factor for SPL 540 mm. The effective FoVs are 6.6$'$, 6.3$'$, 6.0$'$ and 6.3$'$ at 10 keV, 13.5 keV, 30 keV and 60 keV, respectively. All FoVs are less than the collimator's FoVs. As a comparison, the effective FoVs of NuSTAR are around 12$'$ and 6$'$ at 10 keV and 60 keV. (\textbf{e}) The simulated HPD as a function of off-axis angle for SPLs. The HPD remains approximately constant inside the FoV defined by the collimator (9.8$'$ for SPL 109 mm and 6.4$'$ for SPL 540mm).}
     \label{F:offaxis}
\end{figure*}

 \begin{figure*}[h]
 	\centering 
 	\includegraphics[width = 14cm]{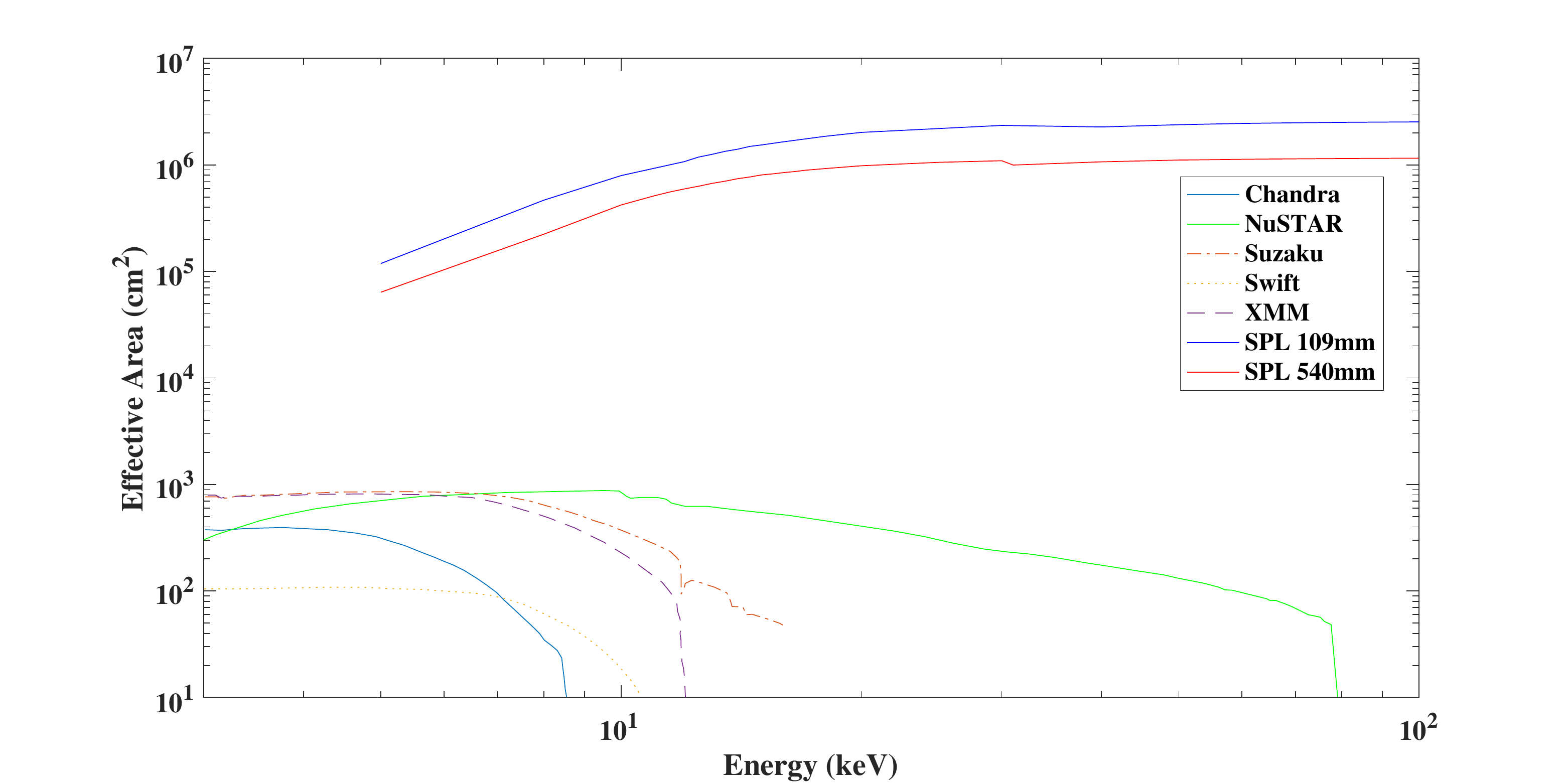}
    \bcaption{A comparison between the potential effective collecting area for an X-ray telescope based on the Stacked Prism Lenses shown in Supplementary Fig.~\ref{F: ASvsenergy}, and selected X-ray telescope missions.}{The effective collecting area is estimated for Falcon Heavy fairing under the same assumption as in Table~\ref{T:para}.}
    \label{F: Evsenergy}
\end{figure*}

 \clearpage
 \begin{figure*}[!p]
 \centering 
 \includegraphics[width = 14cm]{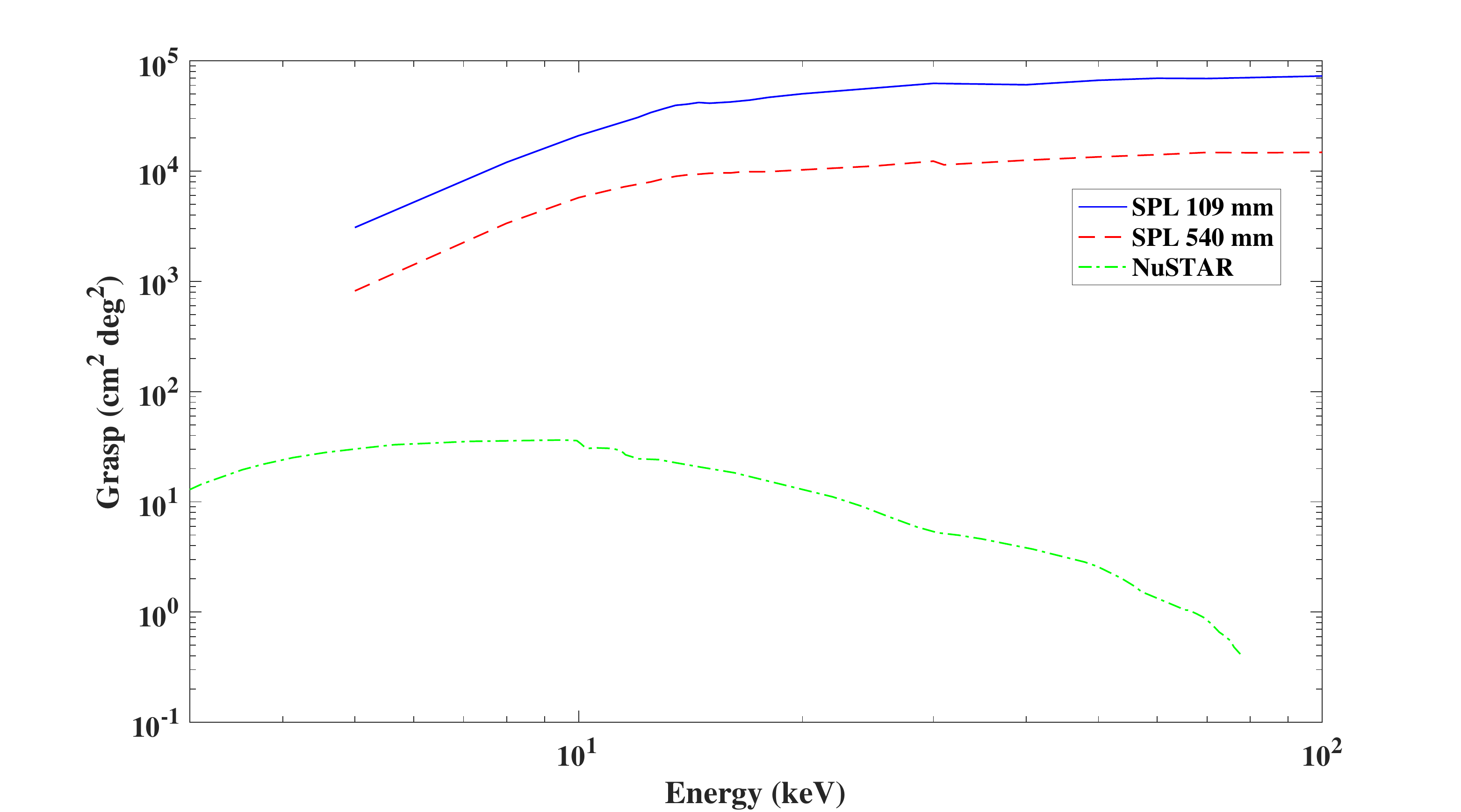}
    \bcaption{Simulated grasp as a function of X-ray energy for an X-ray telescope based on Stacked Prism Lenses, compared to the NuSTAR mission.}{The SPL grasp is estimated for a telescope which half fills the Falcon Heavy fairing.}
    \label{F: grasp}
\end{figure*}
\clearpage

\begin{figure*}[p]
 \centering 
 \includegraphics[width = 12cm]{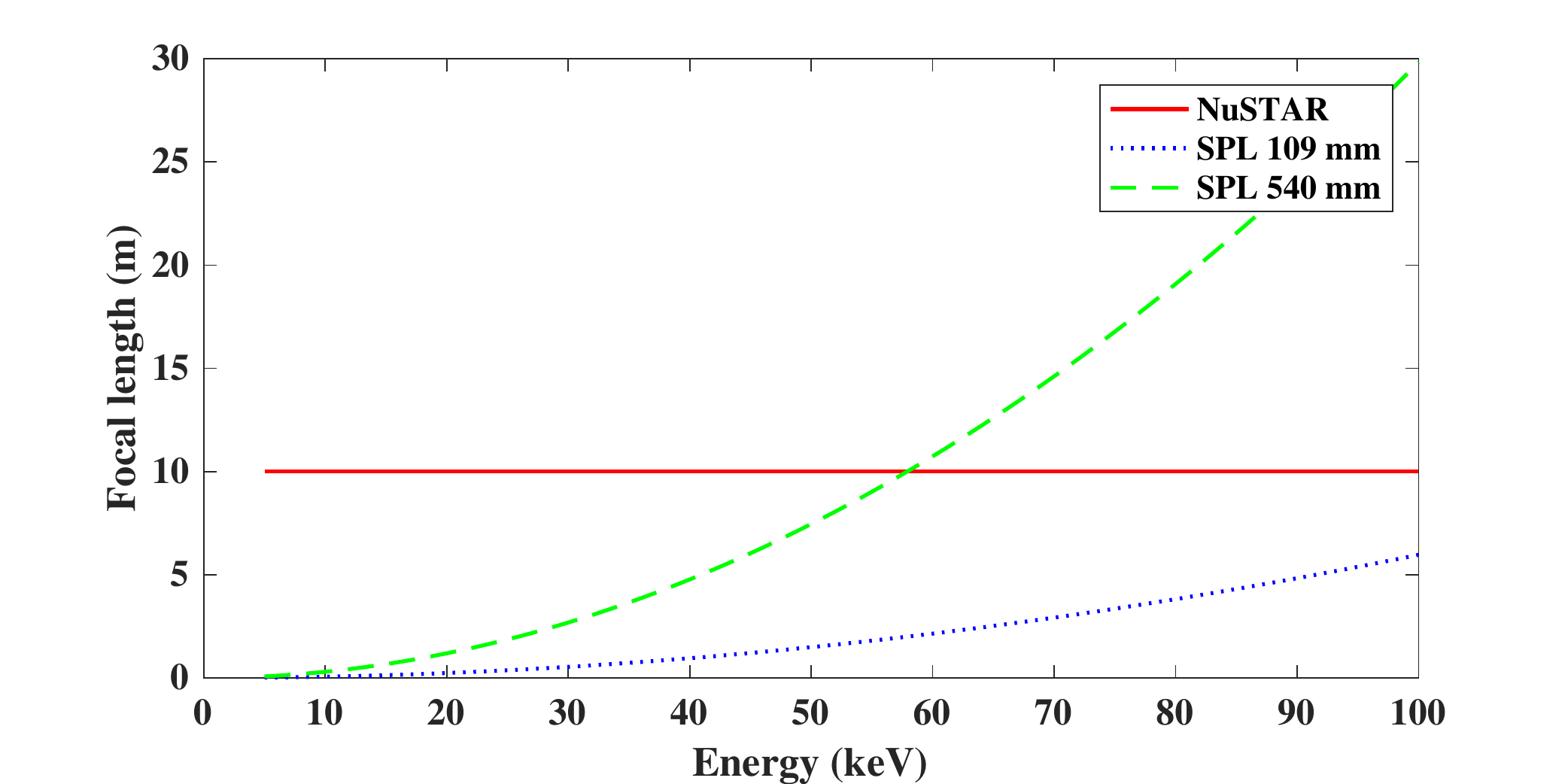}
    \bcaption{The energy dependence of the focal length for Stacked Prism Lenses compared to the NuSTAR mission.} {The Stacked Prism Lens is chromatic since the focal length is proportional to the square of the incident X-ray energy. The focal length for NuSTAR is constant, at around 10 m, across the working energy range.}
    \label{F:focallength}
\end{figure*}

\end{document}